\documentclass[preprint2]{aastex}

\usepackage{epsfig}

\newcommand{\Hubble}{\ensuremath{\mathrm{H}_0}}
\newcommand{\Msun}{\ensuremath{~\mathrm{M}_\odot}}
\newcommand{\Lsun}{\ensuremath{~\mathrm{L}_\odot}}
\newcommand{\LBsun}{\ensuremath{~\mathrm{L}_{B\odot}}}
\newcommand{\AV}{\ensuremath{\mathrm{A}_V}}
\newcommand{\QH}{\ensuremath{Q(\mathrm{H})}}
\newcommand{\BH}{\ensuremath{\mathrm{BH}}}
\newcommand{\MBH}{\ensuremath{M_\mathrm{BH}}}
\newcommand{\Msph}{\ensuremath{M_\mathrm{sph}}}
\newcommand{\Lsph}{\ensuremath{L_\mathrm{sph}}}
\newcommand{\LBsph}{\ensuremath{L_\mathrm{B,sph}}}
\newcommand{\sigstar}{\ensuremath{\sigma_\star}}

\newcommand{\ten}[1]{\ensuremath{10^{#1}}}
\newcommand{\xten}[1]{\ensuremath{\times 10^{#1}}}
\newcommand{\1}{\ensuremath{^{-1}}}
\newcommand{\2}{\ensuremath{^{-2}}}
\newcommand{\3}{\ensuremath{^{-3}}}

\newcommand{\CM}{\ensuremath{\mathrm{~cm}}}
\newcommand{\KM}{\ensuremath{\mathrm{~km}}}
\newcommand{\PC}{\ensuremath{\mathrm{~pc}}}
\newcommand{\KPC}{\ensuremath{\mathrm{~kpc}}}
\newcommand{\MPC}{\ensuremath{\mathrm{~Mpc}}}

\newcommand{\SEC}{\ensuremath{\mathrm{~s}}}
\newcommand{\YR}{\ensuremath{\mathrm{~yr}}}

\newcommand{\kms}{\KM\SEC\1}
\newcommand{\ERG}{\ensuremath{\mathrm{~erg}}}
\newcommand{\ARCSEC}{\ensuremath{\mathrm{~arcsec}}}

\newcommand{\vel}{\ensuremath{\langle v\rangle}}
\newcommand{\vcirc}{\ensuremath{v_\mathrm{circ}}}
\newcommand{\velsq}{\ensuremath{\langle v^2\rangle}}

\newcommand{\HA}{\ensuremath{\mathrm{H}\alpha}}
\newcommand{\HB}{\ensuremath{\mathrm{H}\beta}}
\newcommand{\NII}{\ensuremath{\mathrm{[N\,II]}}}
\newcommand{\OIII}{\ensuremath{\mathrm{[O\,III]}}}

\newcommand{\tst}{\tablenotemark{\ensuremath{\star}}}

\newcommand{\Dx}{\ensuremath{\Delta x}}
\newcommand{\Dy}{\ensuremath{\Delta y}}

\newcommand{\Dw}{\ensuremath{\Delta w}}

\newcommand{\dw}{\ensuremath{\mathrm{d}w}}
\newcommand{\dwp}{\ensuremath{\mathrm{d}w'}}
\newcommand{\xI}{\ensuremath{x}}
\newcommand{\yI}{\ensuremath{y}}
\newcommand{\vI}{\ensuremath{v}}
\newcommand{\wI}{\ensuremath{w}}
\newcommand{\dxI}{\ensuremath{\mathrm{d}x}}
\newcommand{\dyI}{\ensuremath{\mathrm{d}y}}
\newcommand{\dvI}{\ensuremath{\mathrm{d}v}}
\newcommand{\dwI}{\ensuremath{\mathrm{d}w}}
\newcommand{\xII}{\ensuremath{x'}}
\newcommand{\yII}{\ensuremath{y'}}
\newcommand{\dxII}{\ensuremath{\mathrm{d}x'}}
\newcommand{\dyII}{\ensuremath{\mathrm{d}y'}}
\newcommand{\ms}{\ensuremath{m^\prime}}
\newcommand{\mssq}{\ensuremath{m^{\prime 2}}}
\newcommand{\xs}{\ensuremath{x^\prime}}
\newcommand{\xssq}{\ensuremath{x^{\prime 2}}}
\newcommand{\ys}{\ensuremath{y^\prime}}
\newcommand{\yssq}{\ensuremath{y^{\prime 2}}}
\newcommand{\qs}{\ensuremath{q^\prime}}
\newcommand{\qssq}{\ensuremath{q^{\prime 2}}}
\newcommand{\q}{\ensuremath{q}}

\newcommand{\KERN}{\ensuremath{\mathcal{K}}}
\newcommand{\mlr}{\ensuremath{\Upsilon}}
\newcommand{\I}{\ensuremath{i}}
\newcommand{\Th}{\ensuremath{\theta}}
\newcommand{\B}{\ensuremath{b}}
\newcommand{\So}{\ensuremath{\mathrm{s}_\circ}}
\newcommand{\Vsys}{\ensuremath{V_\mathrm{sys}}}
\newcommand{\chisq}{\ensuremath{\chi^2}}
\newcommand{\chisqr}{\ensuremath{\chi^2_\mathrm{red}}}
\newcommand{\chisqc}{\ensuremath{\chi^2_\mathrm{c}}}
\newcommand{\eg}{e.g.}

\slugcomment{Draft version, \today}

\shorttitle{Black Hole at the Center of NGC 4041}
\shortauthors{Marconi et al.}

\begin{document}

\title{
Is there really a Black Hole at the center of
NGC 4041? - Constraints from gas kinematics\altaffilmark{1}}

\author{
	A.~Marconi\altaffilmark{2},
	D.J.~Axon\altaffilmark{3},
	A.~Capetti\altaffilmark{4},
	W.~Maciejewski\altaffilmark{2,11}, 
	J.~Atkinson\altaffilmark{3},
	D.~Batcheldor\altaffilmark{3},
	J.~Binney\altaffilmark{6},
	M.~Carollo\altaffilmark{7},
	L.~Dressel\altaffilmark{5},
	H.~Ford\altaffilmark{8},
	J.~Gerssen\altaffilmark{5},
	M.A.~Hughes\altaffilmark{3}, 
	D.~Macchetto\altaffilmark{5,9},
	M.R.~Merrifield\altaffilmark{10},
	C.~Scarlata\altaffilmark{5},
	W.~Sparks\altaffilmark{5},
	M.~Stiavelli\altaffilmark{5},
	Z.~Tsvetanov\altaffilmark{8},
	R.P.~van~der~Marel\altaffilmark{5} 
}

\altaffiltext{1}{Based on observations made with the NASA/ESA Hubble Space
Telescope, obtained at the Space Telescope Science Institute, which is operated
by the Association of Universities for Research in Astronomy, Inc., under NASA
contract NAS 5-26555. These observations are associated with proposal \#8228.}
\altaffiltext{2}{INAF- Osservatorio Astrofisico di Arcetri,
L.go Fermi 5, I-50125 Firenze, Italy}
\altaffiltext{3}{Department of Physical Sciences, University of Hertfordshire,
Hatfield AL10 9AB, UK}
\altaffiltext{4}{INAF- Osservatorio Astronomico di Torino, Strada Osservatorio 20, 10025 Pino Torinese, Torino, Italy}
\altaffiltext{5}{Space Telescope Science Institute, 3700 San Martin Drive,
Baltimore, MD 21218}
\altaffiltext{6}{Theoretical Physics, 1 Keble Road, Oxford OX1 3NP, UK}
\altaffiltext{7}{Eidgenoessische Technische Hochschule Zuerich, Hoenggerberg HPF G4.3,  CH-8092 Zuerich, Switzerland}
\altaffiltext{8}{Department of Physics and Astronomy, Johns Hopkins University,
3400 North Charles Street, Baltimore, MD 21218}
\altaffiltext{9}{ESA Space Telescopes Division}
\altaffiltext{10}{School of Physics \& Astronomy, University of Nottingham, NG7 2RD, UK}
\altaffiltext{11}{also, Obserwatorium Astronomiczne Uniwersytetu Jagiello{\'n}skiego, Poland}

\begin{abstract}
We present HST/STIS spectra of the Sbc spiral galaxy NGC 4041 which
were used to map the velocity field of the gas in its nuclear region.
We detect the presence of a compact ($r\simeq 0\farcs4\simeq 40\PC$),
high surface brightness, rotating nuclear disk co-spatial with
a nuclear star cluster. The disk is characterized by a rotation curve
with a peak to peak amplitude of $\sim 40\kms$ and is systematically
blueshifted by $\sim 10 - 20\kms$ with respect to the galaxy
systemic velocity.  With the standard assumption of constant
mass-to-light ratio and with the nuclear disk inclination taken
from the outer disk, we find that a dark point mass of
$(1_{-0.7}^{+0.6})\xten{7}\Msun$ is needed to reproduce the observed
rotation curve.  However the observed blueshift suggests the
possibility that the nuclear disk could be dynamically decoupled.
Following this line of reasoning we relax the standard assumptions and
find that the kinematical data can be accounted for by the stellar
mass provided that either the central mass-to-light ratio is
increased by a factor of $\sim 2$ or that the inclination is allowed to
vary. This model results in a $3\sigma$ upper limit of $6
\xten{6}\Msun$ on the mass of any nuclear black hole.  Overall,
our analysis only allows us to set an upper limit of
$2\xten{7}\Msun$ on the mass of the nuclear \BH.  If this upper limit
is taken in conjunction with an estimated bulge B magnitude of $-17.7$
and with a central stellar velocity dispersion of $\simeq 95\kms$,
then these results are not inconsistent with both the \MBH-\Lsph\
and the \MBH-\sigstar\ correlations.
Constraints on \BH\ masses in spiral galaxies of types as late as Sbc are still very scarce and therefore the present result adds an important new datapoint to our understanding of \BH\ demography.
\end{abstract}


\section{Introduction}

It has long been suspected that the most luminous AGN are powered by
accretion of matter onto massive black holes \citep[hereafter \BH;
\eg][]{lind69}.  This belief, combined with the observed evolution of
the space-density of AGN \citep{soltan,chokshi,ms01} and the high
incidence of low luminosity nuclear activity in nearby galaxies
\citep{ho97}, implies that a significant fraction of luminous galaxies
must host black holes of mass $\ten{6}-\ten{10}\Msun$.

It is now clear that a large fraction of hot spheroids (E-S0) contains a \BH\
\citep{harms,kr95,k96,macchetto97,marel98,bower98,marconi01} with mass
proportional to the mass (or luminosity) of the host spheroid ($\MBH/\Msph
\approx 0.001$ \eg\ \citealt{merritt01}).  Recently \citet{ferrarese00} and
\citet{gebhardt00} have shown that a tighter correlation holds between the BH
mass and the velocity dispersion of the bulge.  Clearly, any correlation of
black hole and spheroid properties would have important implications for
theories of galaxy formation in general, and bulge formation in particular.
However, to date, there are very few secure BH measurements or
upper limits in spiral galaxies
even though we know that AGNs are common in such systems \citep{maiolino}.  In
total, there are 37 {\it secure} \BH\ detections 
according to \citet{kg01} or just 22
according to \citet{mf01}, depending on the definition of ``secure''.  
Only $\sim 20\%$ of these \BH\ detections (7/37 or 4/22, respectively) are in
galaxy types later than S0, and only 3 in Sbc's and later
(the Milky Way, \citealt{genzel}; NGC 4258, \citealt{miyoshi}; NGC 4945,
\citealt{greenhill}).
It is therefore important to directly establish how
common are BHs in spiral galaxies and if they follow the same \MBH-\Msph,
\MBH-\sigstar\ correlations as Elliptical galaxies.

This can be achieved only with a comprehensive survey for \BH s that
covers quiescent and active spiral galaxies of all Hubble types. Such
a survey would pin down the mass function and space density of \BH s,
and their connection with host galaxy properties (\eg, bulge mass,
disk mass etc).

To detect \BH s one requires spectral information at the highest
possible angular resolution: the ``sphere of influence'' \citep{rbh} of
\BH s are typically $\le 1$\arcsec\ even in the most nearby
galaxies.  Nuclear absorption-line spectra can be used to demonstrate
the presence of a \BH\ \citep{kr95,r98,marel98b}, but the
interpretation of the data is complex because it involves
stellar-dynamical models that have many degrees of freedom -- that
can be pinned down only when data of very high S/N are
available \citep{binney82,statler,merr97,binney98}.  Radio-frequency
measurements of masers in disks around \BH s provide some of the most
spectacular evidence for \BH s, but have the disadvantage that only a
small fraction of the disks will be inclined such that their
maser emission is directed toward us \citep{braatz}.  Studies of
ordinary optical emission lines from gas disks 
provide an alternative and relatively simple method to
detect \BH s.

HST studies have discovered many cases of such gas disks in early-type
galaxies (M87, \citealt{ford94}; NGC 4261, \citealt{jaffe96}; NGC
5322, \citealt{carollo97}; Cen A, \citealt{schreier98}) and have
demonstrated that both their rotation curves and line profiles are
consistent with thin disks in Keplerian motion
\citep{ferrarese96,macchetto97,ford98,marel98,bower98,marconi01}.

In early-type galaxies there are still worrying issues about the dynamical
configuration of nuclear gas (\eg, misalignment with the major axis, irregular
structure etc).  By contrast nuclear gas in relatively quiescent spirals is
believed  to be organized into well defined rotating disks seen in optical line
images (\eg, M81, \citealt{devereux97}). \citet{ho02} recently
found that the majority of spiral galaxies in their survey has irregular
velocity fields in the nuclear gas, not well suited for kinematical analysis.
Still, $25\%$ of the galaxies where \HA\ emission was detected all the way to
the center have velocity curves consistent with circular rotation and the
galaxies with more complicated velocity curves can also be useful for \BH\ mass
measurement after detailed analysis of the spectra.  Indeed, even in the most
powerful Seyfert nuclei such as NGC 4151, where the gas is known to be
interacting with radio ejecta, it may be possible to get the mass of the \BH\
from spatially resolved HST spectroscopy by careful analysis of the velocity
field to separate the underlying quiescently rotating disk gas from that
disturbed by the jets \citep{winge99}. 

Prompted by these considerations, we have undertaken a spectroscopic
survey of 54 spirals using STIS on the Hubble Space Telescope.  Our
sample was extracted from a comprehensive ground-based study by Axon
et al.\ who obtained \HA\ and \NII\ rotation curves at a
seeing-limited resolution of 1\arcsec, of 128 Sb, SBb, Sc, and SBc
spiral galaxies from RC3. By restricting ourselves to galaxies with
recession velocities $V<2000\,$km/s, we obtained a volume-limited
sample of 54 spirals that are known to have nuclear gas disks and span
wide ranges in bulge mass and concentration.  The systemic
velocity cut-off was chosen so that we can probe close to the nuclei
of these galaxies, and detect even lower-mass black holes.  The frequency of
AGN in our sample is typical of that found in other surveys of nearby
spirals, with comparable numbers of weak nuclear radio sources and
LINERS. The sample is described in detail by Axon et al.\ (paper in
preparation).

This paper presents the observations of NGC 4041, the first 
object observed, and a detailed description of the analysis and modeling
techniques which will be applied to the other galaxies in the sample.
From the Lyon/Meudon Extragalactic Database
(LEDA\footnote{http://leda.univ-lyon1.fr}), NGC 4041 is classified as
a Sbc spiral galaxy with no detected AGN activity.
Its average
heliocentric radial velocity from radio measurements is $1227\pm
9\kms$ becoming $\simeq 1480\kms$ after correction for Local Group
infall onto Virgo. With \Hubble=75\kms\MPC\1\ this corresponds to a
distance of $\simeq 19.5\MPC$ and to a scale of 95\PC/\arcsec.

The outline of the paper is as follows.  In \S\ref{sec:obsdata} we present the
adopted observational strategy and data reduction techniques.  In
\S\ref{sec:results} we present the rotation curves of the ionized gas and the
broad-band images of the nuclear region of the galaxy.  In \S\ref{sec:stardens}
we derive the stellar luminosity density from the observed surface brightness
distribution which is then used in the model fitting of the kinematical data.
All the details of the inversion procedure are described in
Appendix~\ref{app:stardens}.  The model fitting of the kinematical data is
described in \S\ref{sec:modelkin} and all the details of the model computation
are described in Appendix~\ref{app:gaskin}.  In particular,
\S\ref{sec:modstand} describes the {\it standard} approach, while in
\S\ref{sec:modalt} an {\it alternative} approach is considered.  In
\S\ref{sec:discussion} we discuss the effects of our assumptions on the derived
value of the \BH\ mass for which, in the present case, only an upper limit can
be set.  We then compare this upper limit with the \MBH-\Lsph\ and
\MBH-\sigstar\ correlations. Finally, our conclusions are presented in
\S\ref{sec:conclusions}.

\section{\label{sec:obsdata} Observations and Data Reduction}

\subsection{STIS observations}

NGC 4041 was observed with STIS on the HST in 1999 July 7. An $\sim
5\arcsec\times5\arcsec$ acquisition image was obtained with the F28X50LP
filter and the galaxy nucleus, present within the field of view, was
subsequently centered and re-imaged following the ACQ procedure. The exposure
time of the acquisition images was 120\SEC.
\begin{figure}[t!]
\epsfig{figure=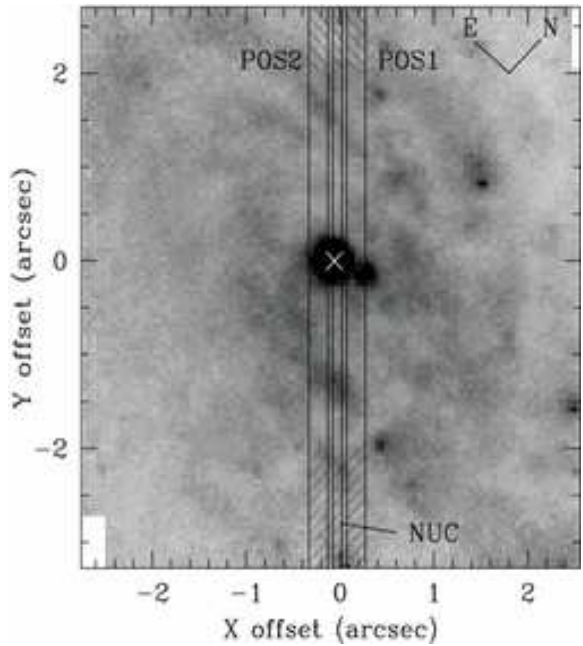,angle=0,width=\linewidth}
\caption{\label{fig:acq} Slit positions overlaid on the acquisition
image. The 0,0 position is the position of the target derived from the
STIS ACQ procedure. The white cross is the kinematic center derived
from the fitting of the rotation curves (see \S\ref{sec:modelkin}).}
\end{figure}

The observational strategy consisted in obtaining spectra at three
parallel positions with the central slit centered on the nucleus and
the flanking ones at a distance of 0\farcs2. The slit positions are
overlaid on the acquisition image in Figure~\ref{fig:acq} and their
position angle is 43\arcdeg.  At each slit position we obtained two
spectra with the G750M grating centered at \HA, with the second
spectrum shifted along the slit by an integer number of detector
pixels in order to remove cosmic-ray hits and hot pixels.  The nuclear
spectrum (NUC) was obtained with the 0\farcs1 slit and no binning of
the detector pixels, yielding a spatial scale of 0\farcs0507/pix along
the slit, a dispersion per pixel of $\Delta\lambda = 0.554$~\AA\ and a
spectral resolution of ${\cal R} = \lambda/(2\Delta\lambda) \simeq
6000$.  The off-nuclear spectra (POS1 and POS2) were obtained with the
0\farcs2 slit and $2\times 2$ on-chip binning of the detector pixels,
yielding 0\farcs101/pix along the slit, 1.108 \AA/pix along the
dispersion direction and ${\cal R}\simeq 3000$.  Total exposure
times were 950\SEC\ for the NUC position and 420\SEC\ and 500\SEC\ for
POS1 and POS2, respectively.

The acquisition images were flat-fielded, realigned and co-added in
order to improve the signal-to-noise ratio. The pixel scale is
$0\farcs0507$.  The flux calibration was first obtained using the
PHOTFLAM header in the image. We subsequently applied a color
correction to convert to Johnson R magnitudes.  In order to do this,
we used the average spectra of Sb and Sc spiral galaxies from
\citet{kinney96} as spectral templates.

The raw spectra were reprocessed through the {\it calstis} pipeline using the
darks obtained daily for STIS. Standard pipeline tasks were used to obtain
flat-field corrected images.  The two exposures taken at a given slit position
were then realigned with a shift along the slit direction (by an integer number
of pixels) and the pipeline task {\it ocrreject} was used to reject cosmic rays
and hot pixels. Subsequent calibration procedures followed the standard
pipeline reduction described in the STIS Instrument Handbook \citep{stishand},
i.e.\  the spectra were wavelength calibrated and corrected for 2D
distortions.  The expected accuracy of the wavelength calibration is
$0.1 - 0.3$ pix within a single exposure and 
$0.2 - 0.5$ pix among different exposures
\citep{stishand} which converts into $\sim 3 - 8\kms$ (relative) and $\sim
5 - 13\kms$ (absolute).  The relative error on the wavelength calibration is
negligible for the data presented here because our analysis is restricted to
the small detector region including \HA\ and \NII\ ($\Delta\lambda<100$\AA).

The nominal slit positions obtained as a result of the STIS ACQ procedure were
checked by matching the light profiles measured along the slit with the
synthetic ones derived from the acquisition image: we collapsed the spectra
along the dispersion direction and compared the resulting light profiles with
the ones extracted from the acquisition image for a given slit position.
The agreement is good for all slit positions and the center of the NUC slit if
offset by only $\sim 0\farcs03$ with respect to the position of the target
determined by the STIS ACQ procedure (0,0 position in Figure~\ref{fig:acq}).

We selected the spectral regions containing the lines of interest and
subtracted the continuum by fitting a linear polynomial row by row along the
dispersion direction.  The continuum subtracted lines were then fitted row by
row with gaussian functions using the task LONGSLIT in the TWODSPEC FIGARO
package \citep{longslit} and the task {\it specfit} in the
IRAF\footnote{IRAF is distributed by the National Optical Astronomy
Observatories, which are operated by the Association of Universities for
Research in Astronomy, Inc., under cooperative agreement with the National
Science Foundation.  } {\it stsdas} package. When the signal-to-noise ratio
(SNR) was insufficient (faint line) the fitting was improved by co-adding two
or more pixels along the slit direction.  

\subsection{WFPC2 images}

WFPC2 images in the F450W ($\sim$B), F606W ($\sim$R) and F814W
($\sim$I) filters were retrieved from the archive. 
These images encompass the entire galaxy with the nucleus
located in the WF3 chip. The data were
automatically reprocessed with the best calibration files available
before retrieval.  Two exposures were performed in each of the three
filters, in order to remove cosmic rays.  Warm pixels and cosmic rays
were removed (STSDAS tasks {\it warmpix} and {\it crrej}) and
mosaicked images with spatial sampling of 0\farcs1/pix were obtained
using the {\it wmosaic} task, which also corrects for the optical
distortions in the four WFPC2 chips. The background was estimated from
areas external to the galaxy where no emission is detected.  Flux
calibration to Vega magnitudes was performed using the zero points by
\citet{wfpc2cal}. To convert to standard filters in the
Johnson-Cousins system we estimated the color correction using various
spectral templates, A0V and K0V stars and the Sb and Sc spiral spectra
from \citet{kinney96}.  The color corrections are the following:
$I-F814W\simeq -0.1$ (Sb, K0V) and $-0.005$ (A0V);
$R-F606W\simeq -0.4$ (Sb, K0V) and $-0.05$ (Sc,A0V);
$B-F450W\simeq 0.1$ (Sb, K0V) and 0.0 (Sc,A0V).
With these corrections, the differences between colors from
Johnson-Cousins and HST instrumental magnitudes are
$\Delta(R-I)\simeq -0.3; -0.04$, $\Delta(B-I)\simeq 0.2; 0.0$,
$\Delta(B-R)\simeq 0.5; 0.0$ according to the templates used.  Unless
stated otherwise, we have applied the color corrections derived from
the K0V star and Sb spiral templates, which are most suitable to the
present data.
\begin{figure*}[t!]
\centerline{
 \epsfig{figure=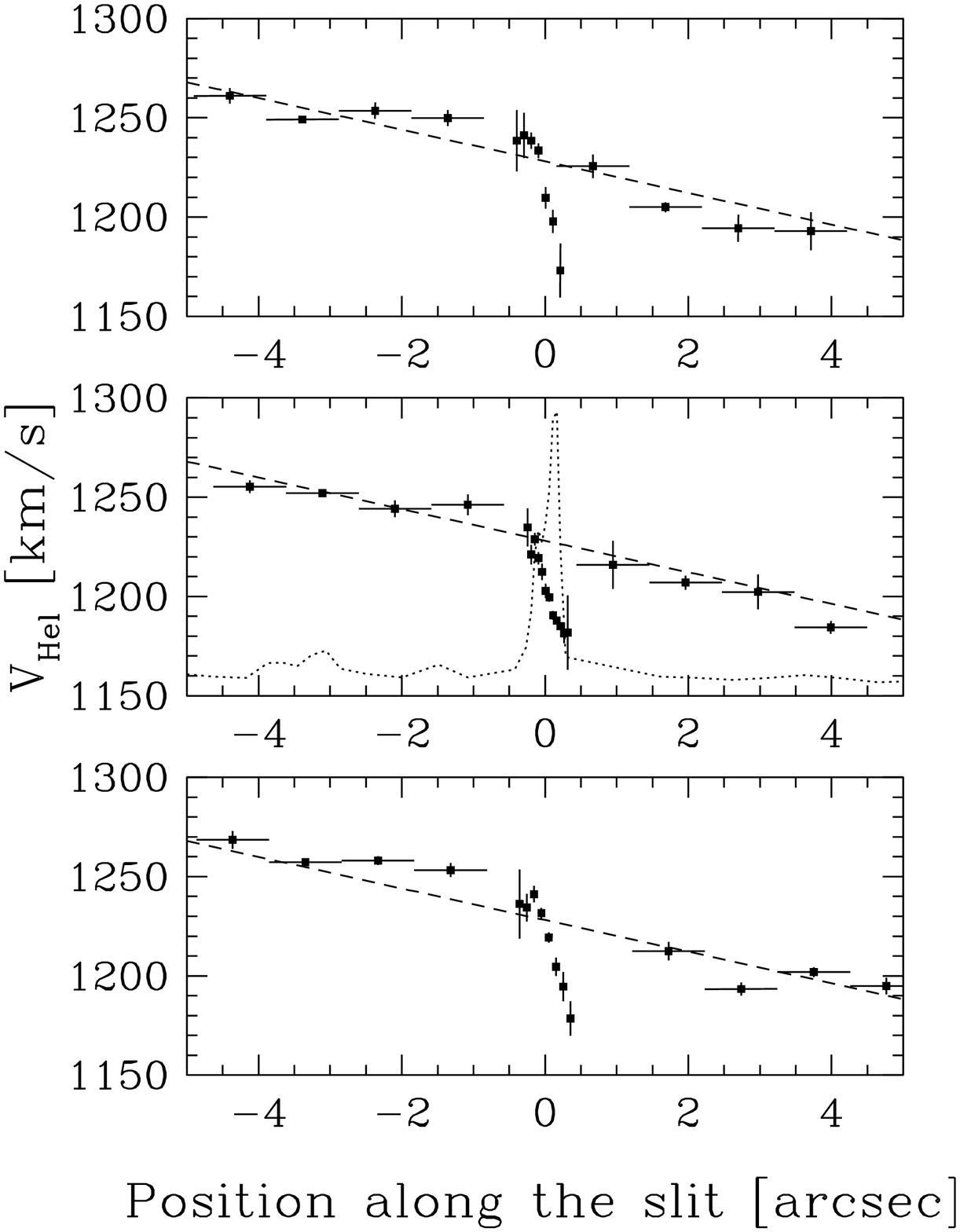,angle=0,width=0.33\linewidth}
 \epsfig{figure=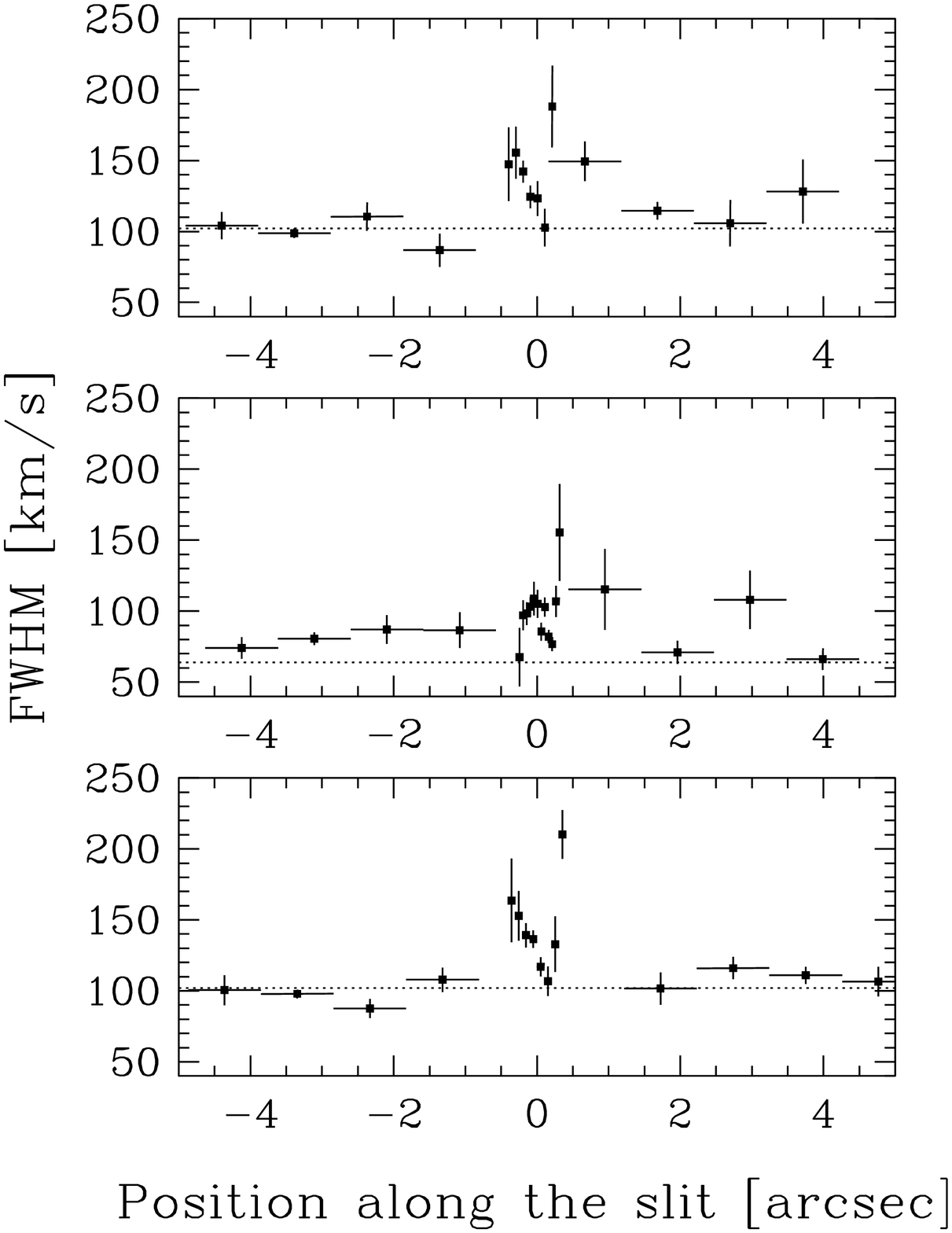,angle=0,width=0.33\linewidth}
 \epsfig{figure=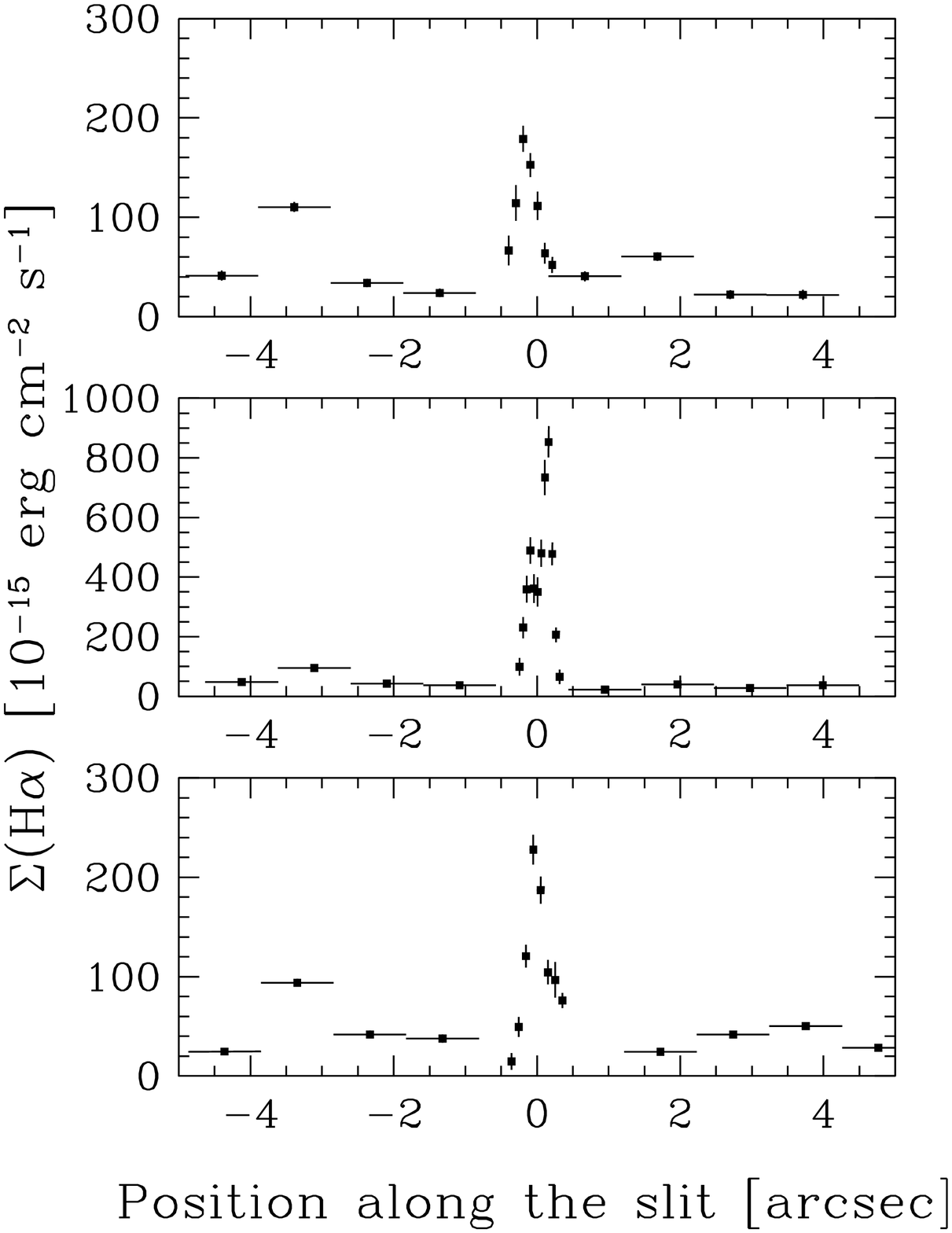,angle=0,width=0.33\linewidth}
}
\caption{\label{fig:rotlarge} From left to right: velocity, FWHM and Surface
brightness measured along the slit at POS1, NUC and POS2 (from top to bottom).
Vertical bars are 1$\sigma$ errors while horizontal bars indicate the size of
the aperture over which the quantity was measured.  In the left panel, the
dashed line is the velocity gradient measured in the ground based data. The
dotted line in the NUC rotation curve panel is the line surface brightness
along the slit, drawn in order to pinpoint the high surface brightness nuclear
disk. The dotted lines in the FWHM panels corresponds to instrumental widths.
The 0 position along the slit corresponds to the position of the nuclear
continuum peak. }
\end{figure*}
\begin{figure*}[t!]
\centerline{
 \epsfig{figure=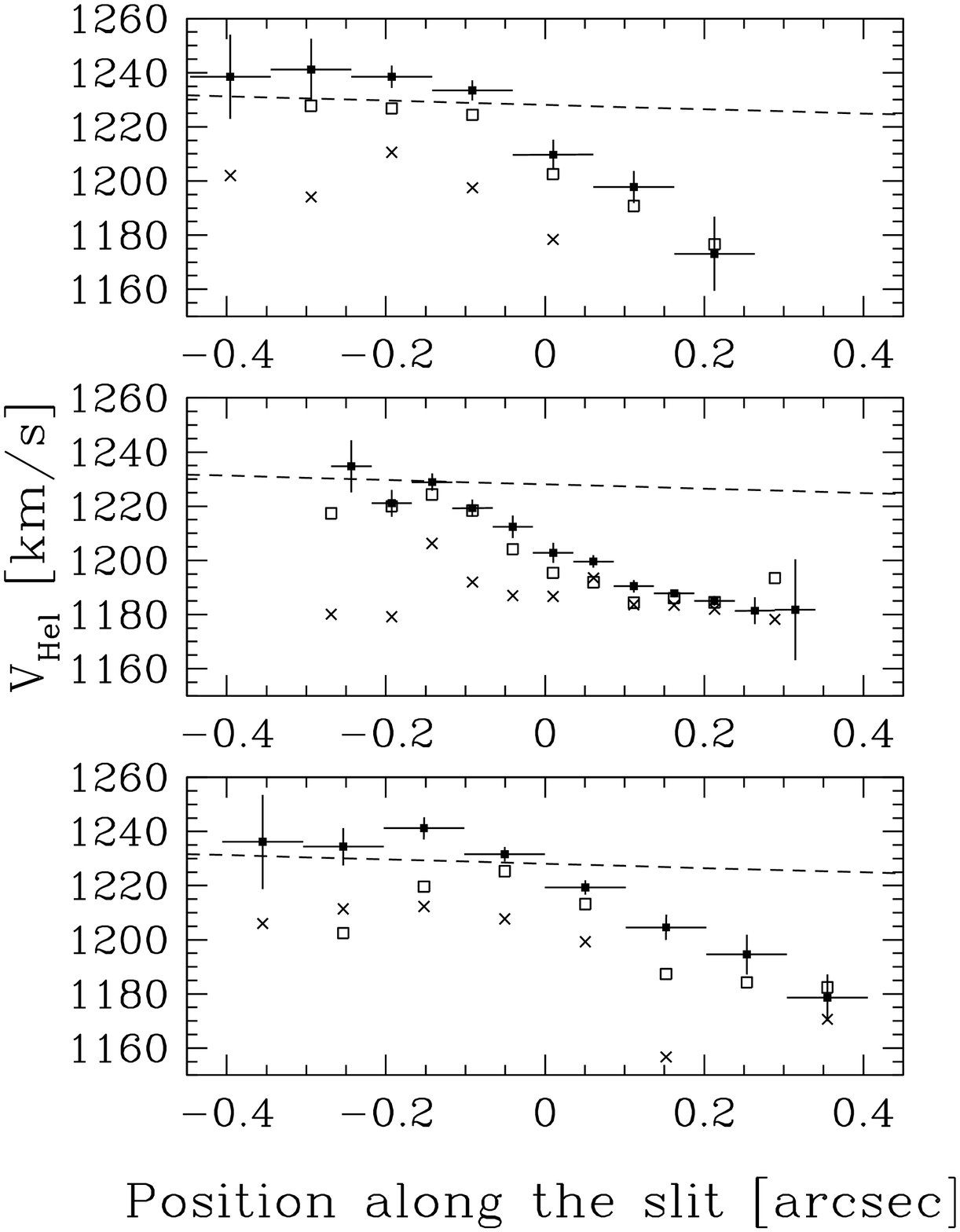,angle=0,width=0.33\linewidth}
 \epsfig{figure=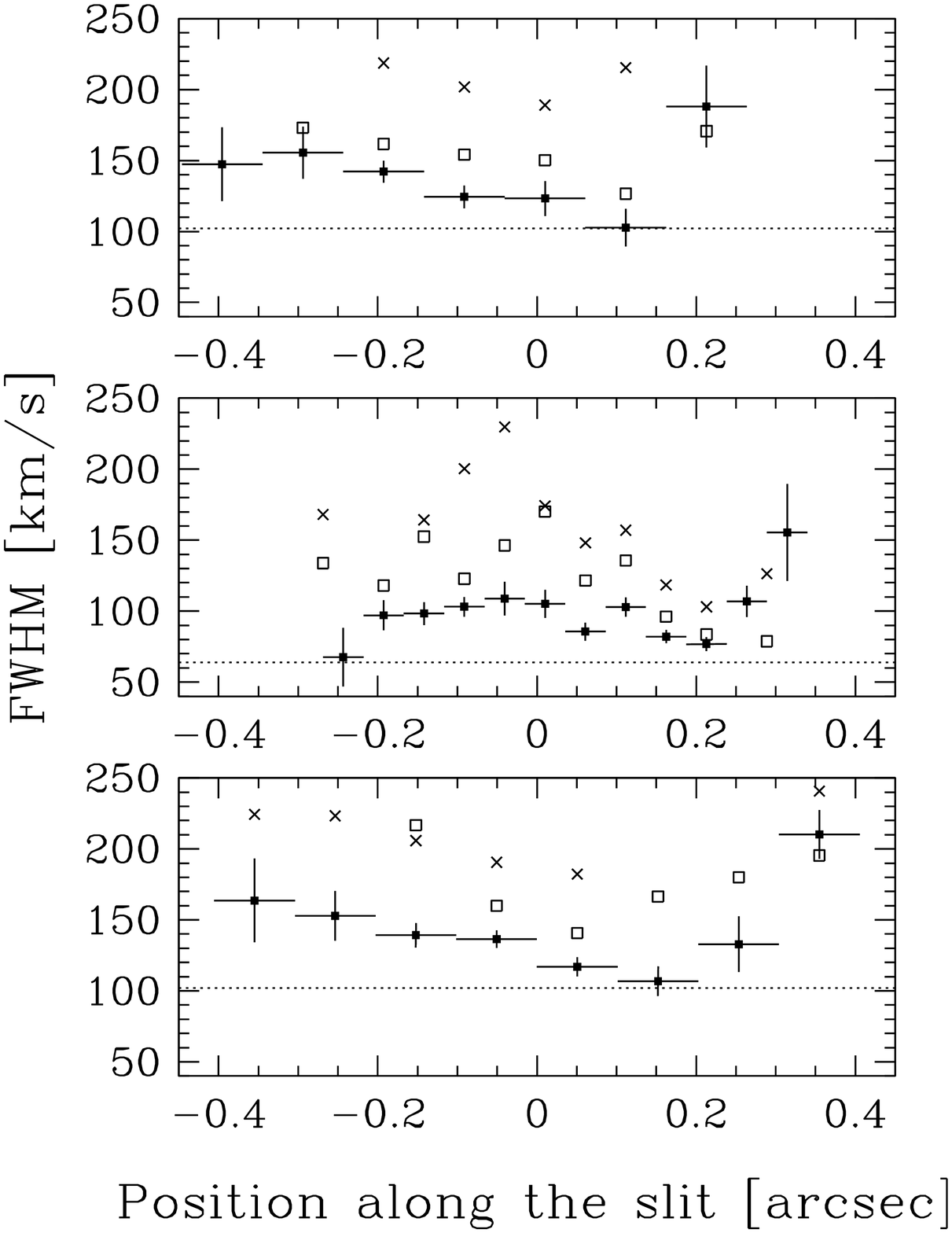,angle=0,width=0.33\linewidth}
 \epsfig{figure=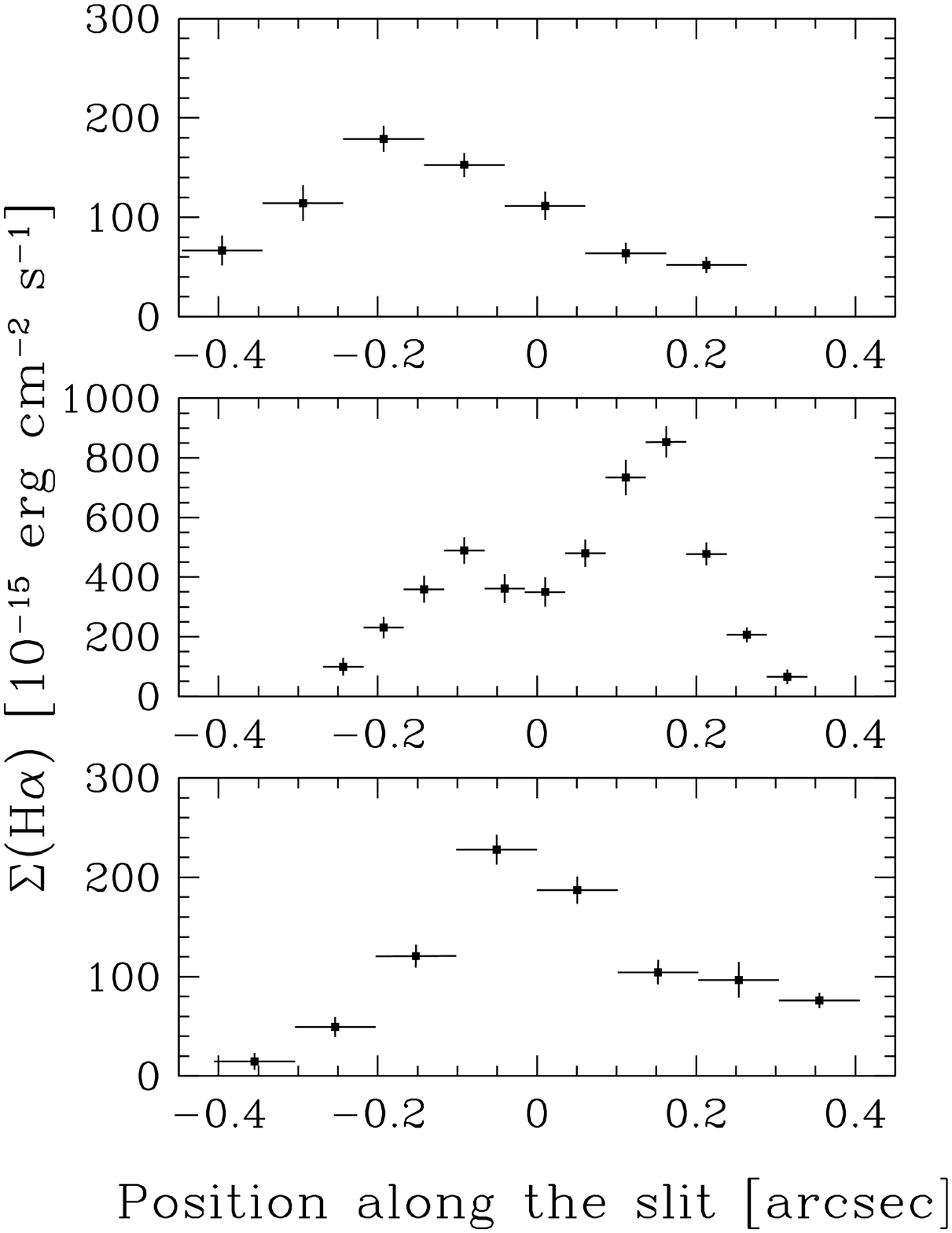,angle=0,width=0.33\linewidth}
}
\caption{\label{fig:rotnuc} Same as previous figure but for the points related
to the nuclear disk. The points with the errorbars are the quantities derived
in the fit which takes into account the presence of the blue wing and in which
\HA\ and \NII\ emitting media are constrained to have the same velocity and
FWHM.  Conversely the points without the errorbars are derived with
unconstrained, single gaussian fits of \HA\ (open square) and \NII\ (crosses).
For more details see \S\ref{sec:linefitting}. }
\end{figure*}
\begin{figure*}[t!]
\epsfig{figure=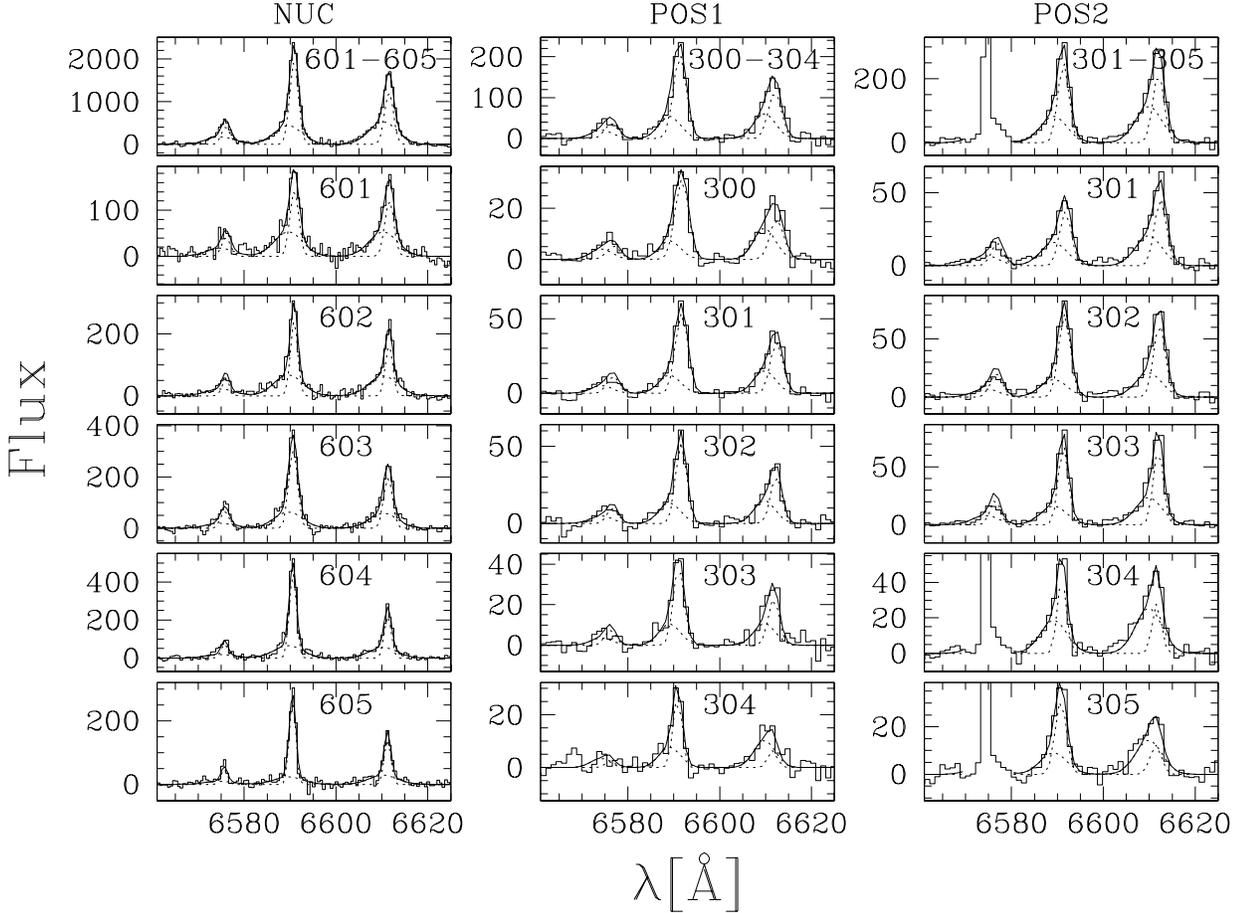,angle=-90,width=\linewidth}
\caption{\label{fig:fits} Fits of the line profiles at the three slit
positions. The dotted lines identify the two components of the fit.  Each
component is characterized by the same velocity and FWHM for \HA\ and \NII. The
ratio between the two \NII\ lines is that fixed by atomic physics.  At every
slit position the blue component has the same velocity and width at each row
and their values are determined in the fit of the overall nuclear spectrum
shown in the upper panels. The numbers in the upper right corners of each panel
are the row or the range of rows where the fit was performed.  The strong
narrow lines in the right panels are residual cosmic rays 
which have been excluded from the fit.}
\end{figure*}

\subsection{Ground based observations}

NGC 4041 was also observed with NICS \citep{nics} at the Telescopio
Nazionale Galileo (TNG) in 2001 February 12 using the $K^\prime$
filter and the small field camera which yields a 0\farcs13 pixel
size. The night was not photometric and seeing during the observations was
$\sim 0\farcs8$. Observations consisted of several exposures, each
with the object at a different position on the array.  Data reduction
consisted in flat-fielding and sky subtraction. The frames were then
combined into a mosaic.  In order to flux calibrate the NICS image we
have used the 2MASS image available on the web. The 2MASS K band
image is flux calibrated to an accuracy of $<0.1$mag and has a spatial
resolution of $\sim 3\farcs5$ with a pixel size of 1\arcsec.  We have
therefore degraded the spatial resolution of the NICS image to the
2MASS image and rebinned to 1\arcsec\ pixels. We have performed
ellipse fitting to both images and rescaled the NICS image to the flux
calibrated 2MASS data by comparing the light profiles.

The NICS image and WFPC2 images were then realigned by cross
correlating common features in the nuclear region, because no point
sources are present in the near infrared image.  The main feature
which drives the correlation is the strong nuclear peak, however we
have checked that, even excluding the central peak, the shift between
the images can be determined with an accuracy of $\pm 0\farcs2$
(i.e.\ $\pm 2$ pixels of WFPC2) in both directions. With this accuracy
the nuclear continuum peaks present in both optical and infrared
images are consistent with being at the same position.

\section{Results\label{sec:results}}

\subsection{Kinematics}

\subsubsection{\label{sec:linefitting}Line fitting procedures}

Line-of-sight velocities, full width at half maxima (hereafter FWHM)
and surface brightnesses along each slit were obtained by fitting
single gaussians to \HA\ and \NII\ emission lines in each row of the
continuum-subtracted 2D spectra.  In the left panels of
Figure~\ref{fig:rotlarge} we plot the measured velocities.  Beyond
$0\farcs5$ of the nucleus velocities show considerable small scale
variations likely due to local gas motions which do not reflect the
mass distribution.  Therefore we averaged velocities by binning the
spectra in steps of 1\arcsec\ (10 rows in NUC and 5 in POS1,2), while,
within $0\farcs5$ of the nucleus velocities are measured along each
row to take advantage of all the spatial information.  Similarly, the
central and right panels in Figure~\ref{fig:rotlarge} display FWHMs and
\HA\ surface brightnesses as measured along the slit.
\begin{figure}[t!]
\epsfig{figure=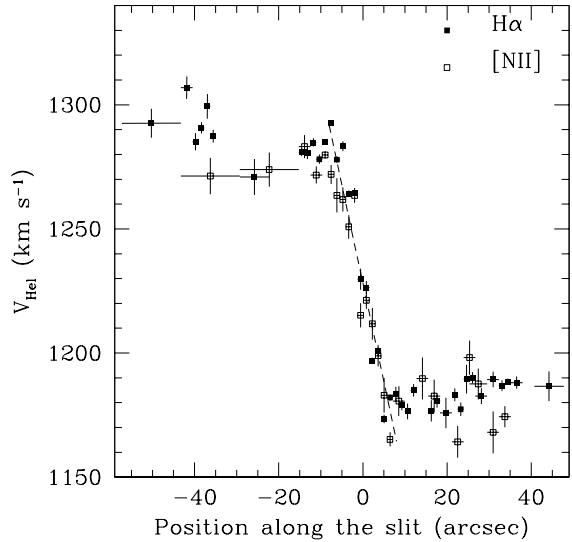,angle=0,width=\linewidth}
\caption{\label{fig:ground} Ground based rotation curve obtained at
PA=43\arcdeg.
The dashed line is the constant velocity gradient from the solid body part of
the rotation curve also shown in Figures~\ref{fig:rotlarge} and \ref{fig:rotnuc}.
} 
\end{figure}

In Figure~\ref{fig:rotlarge}, the dotted line superimposed on the NUC
rotation curve represents the \HA\ surface brightness shown in the
right central panel.  This helps to distinguish the presence of two
components: an extended one, characterized by a low surface brightness
($<10^{-13}\ERG\CM\2\SEC\1\ARCSEC\2$),
roughly constant along the slit, and nuclear one,
compact (within $-0\farcs4$ and 0\farcs4), bright and cospatial with the
position of the nuclear continuum peak.  This might be interpreted as
a nuclear disk of the same extent as the nuclear stellar cluster
described in \S\ref{sec:morpho}.  As shown in more detail in the right
hand panel of Figure~\ref{fig:rotnuc}, the emission line surface
brightness of the nuclear disk is double peaked while the nuclear continuum
source, roughly coincident with the center of rotation, is located 
in--between the two peaks.

Fitting single unconstrained gaussians is acceptable for the extended
component but it does not produce good results for the points in the
nuclear region.  In particular a single gaussian fit produces
velocities of \HA\ and \NII\ which differ by as much as $\sim
30 - 40\kms$ (see Figure~\ref{fig:rotnuc}).  This is not a worrisome issue
if the amplitude of the rotation curve is a few 100 \kms\ but makes
interpretation of the data uncertain in the present case where the
amplitude of the nuclear rotation curve is only $\sim 40\kms$. A
careful analysis of the line profiles in the nuclear region shows that
they are persistently asymmetric with the presence of a blue wing (see
Figure~\ref{fig:fits}).

A fit row by row with two gaussian components, the main component and the blue
one, with the constraint that they have the same velocities and widths for \HA\
and \NII, shows that, within the large uncertainties, the ``blue wing'' has
always the same velocity and width.  We have then deblended the ``blue''
component in the spectrum obtained by co-adding the central 5 rows.
The 
velocity and width of the blue component were then used in the row by row
fit. The constrained fit is good and \HA\ and \NII\ now have the same velocity
in the main component. The measured velocities, FWHMs and \HA\
surface brightnesses in the nuclear region are shown 
in Figure~\ref{fig:rotnuc} where we also plot, as a comparison, the values
obtained from unconstrained single gaussian fits of \HA\ and \NII.
Given the signal-to-noise of the present data, the nature of the blue wing
is, as yet, unclear.
\begin{figure*}[t!]
\epsfig{figure=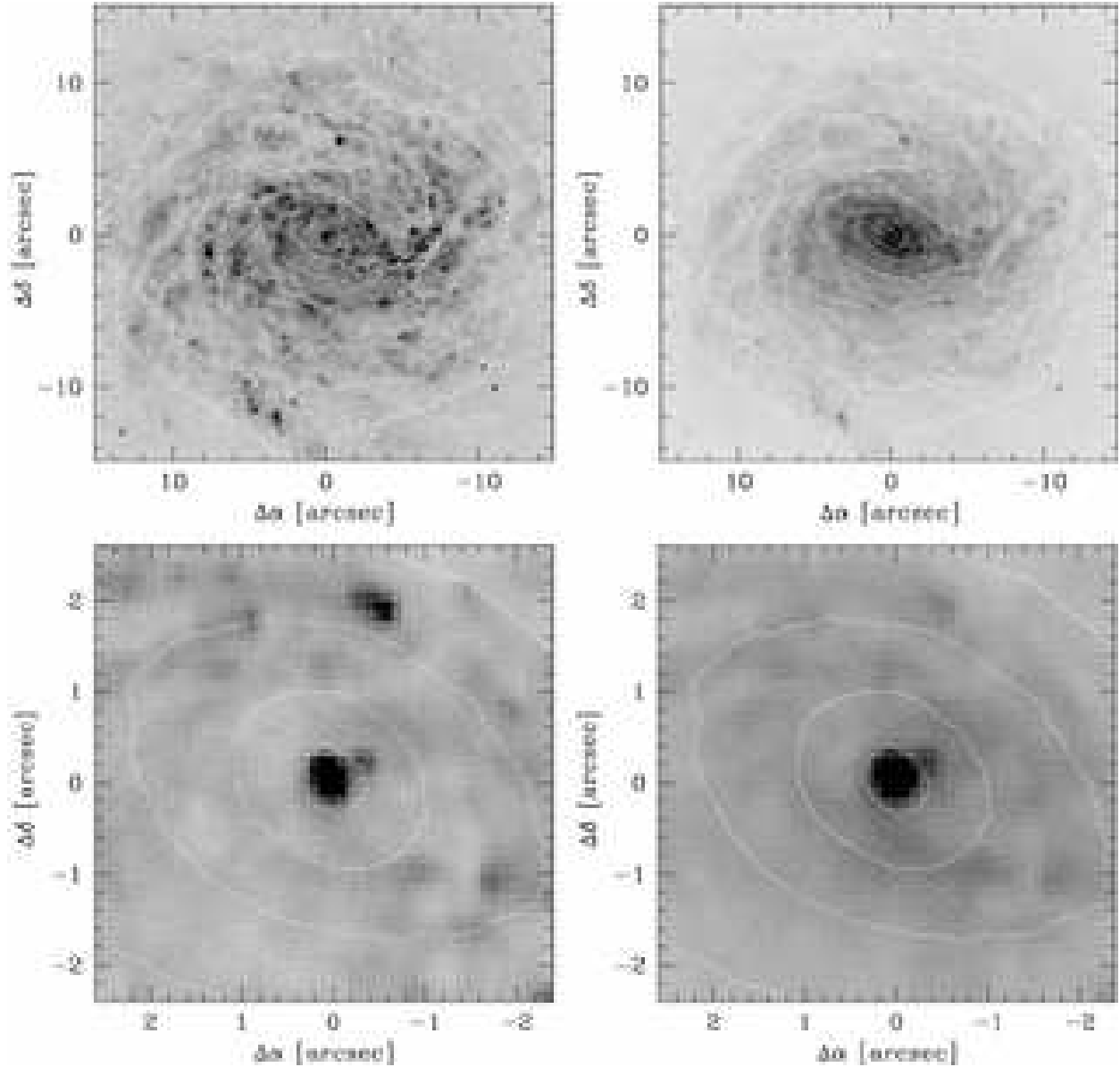,angle=0,width=\linewidth}
\caption{\label{fig:images}Overlay of the K band isophotes on the
grayscales of the F450W (left) and F814W (right) WFPC2 images.
North is up and East is left. The bottom panels show an expanded view
of the nuclear region.}
\end{figure*}
\begin{figure*}[t!]
\epsfig{figure=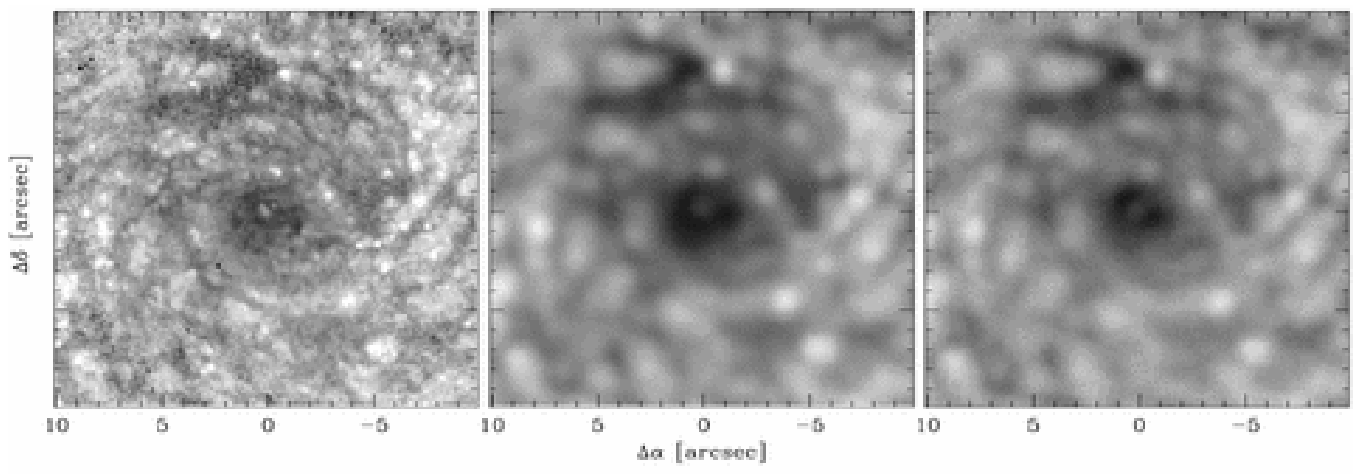,angle=-90,width=\linewidth}
\caption{\label{fig:colours}$20\arcsec\times 20\arcsec$
view of the color images of the central region of
NGC 4041. The center of the images coincides with the location of the
nucleus. North is up and East is left. From left to right: $B-I$,
$B-I$ degraded to NICS resolution, $B-K$ with B (from WFPC2/F450W)
degraded to NICS resolution. The dark regions have redder colors.
}
\end{figure*}

\subsubsection{\label{sec:vcurves}Velocity curves}

The velocity field of the extended component shows a quasi-linear
gradient of $\sim 8 \kms/\arcsec$ which agrees with the velocity
gradient (Figure~\ref{fig:ground}) measured from ground-based observations 
at the same PA as our STIS spectra (Axon et al., paper in preparation).
The velocity gradient measured from the ground based
rotation curve is shown as a straight line in the figures. The large
scale trend observed from HST matches very well the expected gradient
from the solid body part of the ground based rotation curve although it
presents structures at small scales.

Within 0\farcs5 of the nucleus, the velocity field shows a smooth
S-shaped curve with a peak to peak amplitude of only 40\kms\ 
(left panel of Figure~\ref{fig:rotnuc}).  Overall
there is no hint of a steep Keplerian rise around a point-like mass.

The central velocities of the nuclear curves are systematically offset
from the large scale velocity field by $\sim 10 - 20 \kms$ (compare with
the dashed line which is the solid body part of the rotation curve
- left panel of Figure~\ref{fig:rotnuc}).  The off-nuclear slits show essentially
the same velocity field as the on-nucleus slit. Thus this blueshift
must be real and not an instrumental artifact generated by light
entering the slit off center (for a detailed discussion on the effects
of light entering the slit off-center see \citealt{witold01}).
For example, consider the case in which the peak of line emission
is on the blue side (left on Figure~\ref{fig:acq}) of the NUC slit. Then
the measured velocity will be blueshifted with respect to the true
value. The same would happen for POS1.  However in POS2 the peak of
the line emission will be on the red side of the slit and the measured
velocities will be shifted to the red with respect to the true value.
This is not the case for our data.  The possible origin and
implications of this blueshift are discussed in
\S\ref{sec:discussion}.

\subsection{\label{sec:morpho}Morphology}

The acquisition image, shown in Figure~\ref{fig:acq}, has a
field-of-view of $5\arcsec\times5\arcsec$. In Figure~\ref{fig:images} we
show the inner $30\arcsec\times30\arcsec$ and $5\arcsec\times5\arcsec$
of the WFPC2 and NICS images. Finally in Figure~\ref{fig:colours} we
plot the color maps.

\begin{figure}[t!]
\epsfig{figure=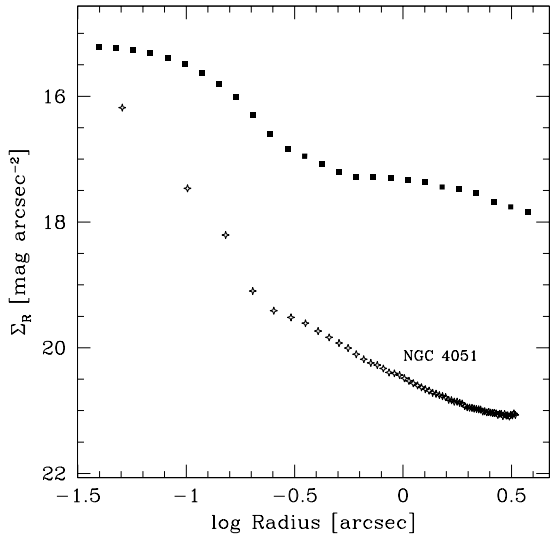,angle=0,width=\linewidth}
\caption{\label{fig:acqprof} Radial light profile from the STIS acquisition
image. The empty diamonds represent the radial light profile from a galaxy, NGC
4051, having a strong central unresolved source; the profile was rescaled to
match the NGC 4041 one in the central pixel. } 
\end{figure}

The acquisition image (Figure~\ref{fig:acq}) shows a compact but
resolved (FWHM$\simeq0\farcs2$) bright central feature superimposed to
the central region of the bulge.  A second bright feature is present
$\sim 0\farcs4$ SW of the nucleus.  Figure~\ref{fig:acqprof} represents
the radial light profile from the STIS acquisition image.  As already
clear from the acquisition image, the light profile indicates the
presence of a bright central feature. This feature is spatially
extended as can be seen from a comparison with the light profile of an
unresolved source (NGC 4051) obtained with the same instrumental
setup.  This nuclear feature is likely to be a star cluster; a
photometrically distinct star cluster is often present in the
dynamical center of spiral galaxies of all Hubble types
\citep[\eg][and references therein]{carollo98,carollo02,boker02}.

The same central source is visible in all the WFPC2 images and an
analysis of its colors (Figure~\ref{fig:colours}) indicates that it is
bluer than the surrounding regions but still redder than the galaxy
disk and bulge.  
Its Vega magnitudes in the three WFPC2 filters are 19.10 (F450W),
18.05 (F606W) and 17.16 (F814W) with formal uncertainties of $\pm
0.02$mag.  These values, converted to Johnson magnitudes, become
B=19.2, R=17.7, I=17.1 for the K0V and Sb spiral spectra and B=19.1,
R=18.0, I=17.15 for the A0V and Sc spiral spectra.  The star cluster
emission is very weak in the K band and not readily visible but its
location is coincident with the location of the K band peak within the
alignment uncertainties ($\sim 0\farcs2$, see Figure~\ref{fig:images}).

The color images show that the inner few arcseconds are redder
than the galaxy disk and bulge and this could either be due to the
presence of obscuration by dust or to a change in stellar population.

\begin{figure}[t!]
\epsfig{figure=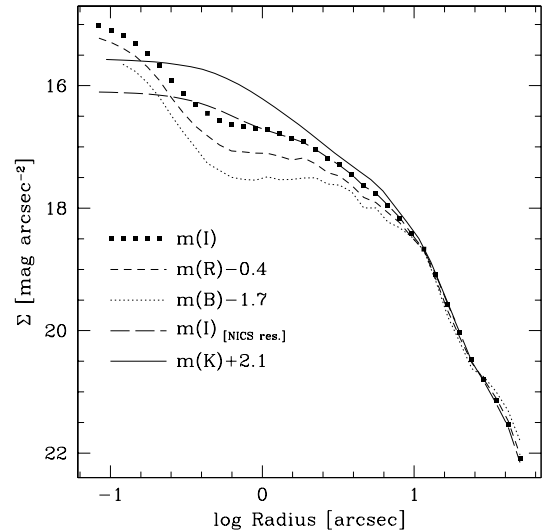,angle=0,width=\linewidth}
\caption{\label{fig:profiles} Azimuthally averaged light profiles for the WFPC2
and NICS images. All images have been rescaled to match the external light
profile of the F814W image. The scaling factors applied are indicated in the
figure and directly show the color in the outer regions where the profiles
match. ``NICS res.'' indicates that the F814W image has been degraded
to the NICS resolution.} 
\end{figure}

From several galaxy catalogues it can be inferred that the inclination
of the large scale disk is $\sim 20$\arcdeg. This is supported by the
nearly circular isophotes observed in the NICS and WFPC2 images at
large scales $> 20\arcsec-30\arcsec$. However, in the inner regions, a
bar-like structure is visible in the K band image oriented $\sim$ E-W.
It is extended for $\sim 15\arcsec$ i.e.\ $\sim 1.4\KPC$ with isophotes
symmetrically twisted on both ends. The twisting is most likely due to
spiral arms just outside (and apparently connecting to) the bar. These
give the bar the appearance of being larger than it is.  It is well
known \citep[see, \eg][]{shaw95,knapen} that strong concentrations of
young stars can dominate the K-band. This likely happens in the region
where spiral arms emanate from the bar in NGC~4041: blue-star
complexes are seen in color images there, particularly at the west end
(Figure~\ref{fig:images}). The bar itself is rather short with semi-major
axis length at most $\sim 3\arcsec$, and a position angle of about 60\arcdeg.
It is a weak bar, given the fatness of the observed
structure and that it is lacking the pair of dust lanes characteristic for
strong bars \citep{athanassoula92}.  Instead, a mini-spiral structure
is seen in the STIS ACQ image (Figure~\ref{fig:acq}), extending at least
out to $3\arcsec$ from the nucleus, i.e.\ throughout the entire IR bar.
Such a morphology indicates that we shouldn't expect strong departures
from the circular motion in the gas flow. In fact, the CO velocity
field observed by \citet[see their Figure~1f]{sakamoto99} shows no such
deviations down to the $4\arcsec$ beam size of the observations. The
PA of the kinematical line of nodes is~$\simeq 85$\arcdeg, close to the PA
of the bar.

\subsubsection{\label{sec:profiles}Light Profiles}

The canonical way to derive light profiles from images is to fit
ellipses to the isophotes and to measure encircled fluxes.  However,
inspection of Figures~\ref{fig:acq} and \ref{fig:images} indicates the
presence of small scale structures invalidating the ellipse fits of
the WFPC2 and STIS images.  The NICS image is much smoother but the
ellipticity of its inner isophotes is caused by the presence of a bar
and spiral arms and not by geometrical projection effects.  Since the
inclination of the large scale disk is only $\simeq 20$\arcdeg, one
expects almost circular isophotes, and it is reasonable to obtain the
light profiles from azimuthal averages instead of canonical ellipse
fitting.

In Figure~\ref{fig:profiles} we plot the azimuthally averaged light
profiles extracted from the WFPC2 and NICS images.  The light profiles
have all been rescaled to match the external one of the F814W
image. The profiles all match well beyond $r>10\arcsec$ and the colors
of that region can be estimated as $B-I\sim 1.7$, $R-I\sim 0.4$ and
$I-K\sim 2.1$.  These colors are consistent with other spiral galaxies:
for instance \citet{dejong96} has shown that the spirals of this
Hubble type on average have integrated colors $B-I=1.8\pm 0.2$
(observed 1.7) and $B-K=3.5\pm 0.3$ (3.8).

The red feature observed in the color maps around the nucleus results
in a flattening of the light profiles which increases at
decreasing wavelength.
From the figure one can immediately estimate the color excess
or reddening for each pair of bands with respect to the external points.
Roughly $E(B-I)\sim 0.8$, $E(R-I)\sim 0.4$ and $E(I-K)\sim 0.5$.  Assuming the
galactic extinction law by \citet{cardelli89} one finds that
with $\AV\sim 1.2$ mag one can approximately match the observed color excesses
($\AV\sim 1.2$ implies $E(B-I)\sim 1.03$, $E(R-I)\sim 0.32$ and $E(I-K)\sim
0.44$).  Thus the red feature observed in the color maps is likely to be due
to extinction by dust with average value of $\AV\sim 1.2$ mag. 

\section{Model fitting}
\begin{deluxetable}{ccccccccc}
\tablewidth{0pt}
\tablecolumns{9}
\tablecaption{\label{tab:fitprofs}
Best fit parameters of the stellar light density.}
\tablehead{ \colhead{$q$} &
\colhead{$\rho^\star_b$\,\tablenotemark{a}} & \colhead{$r^\star_b$\,\tablenotemark{b}} & \colhead{$\beta^\star$}
& \colhead{$\rho_b$\,\tablenotemark{a}} &
\colhead{$m_b$\,\tablenotemark{b}} & \colhead{$\alpha$} & \colhead{$\beta$} &
\colhead{\chisqr} }
\startdata
\cutinhead{I band (WFPC2/F814W)}
1.0 & 1023 & 0.27 & 4.2 & 0.73 & 12.6 & 0.9 & 1.6 &  10.8 \\
0.1 & 951  & 0.31 & 5.0 & 5.8  & 12.6 & 0.9 & 1.6 &  10.5 \\
\cutinhead{K band (NICS)}
1.0 & 910  & 0.27\tablenotemark{c} & 4.2\tablenotemark{c} & 0.44 & 28.0 & 1.4 & 3.0 & 9.8 \\
0.1 & 609  & 0.31\tablenotemark{c} & 5.0\tablenotemark{c} & 1.60 & 45.9 & 1.4 & 6.6 & 7.7 \\
\enddata
\tablenotetext{a}{In units of \Msun\PC\3\ assuming $\mlr=1$.}
\tablenotetext{b}{In units of arcsec.}
\tablenotetext{c}{Kept fixed at the value from the previous fit.}
\end{deluxetable}

\subsection{\label{sec:stardens}The stellar luminosity density}

In order to estimate the stellar contribution to the gravitational
potential in the nuclear region, we have to derive the stellar
luminosity density from the observed surface brightness distribution.
This inversion is not unique if the gravitational potential does not
have a spherical symmetry.  Here, we assume that the gravitational
potential is an oblate spheroid \citep[\eg,][]{marel98}.
We further assume that the principal plane of the potential has the
same inclination as the large scale disk ($\I\simeq20$, LEDA and
\citealt{rc3}).  The knowledge of \I, combined with the observed axial
ratio of the isophotes, should directly provide the axial ratio of an
axisymmetric light distribution, but the presence of a bar-like
structure invalidates this approach since the flattening of the
isophotes cannot be ascribed to simple projection effects.  Therefore
we consider two extreme cases: a spherical and a disk light
distribution.  The model light profiles are computed taking into
account the finite spatial resolution and pixel size of the
detector. A detailed description of the relevant formulas for the
inversion and fitting procedure is presented in Appendix~\ref{app:stardens}.

The light profiles derived from HST images have better spatial
resolution than the one derived from ground-based NICS image. However,
as shown above, the optical HST images suffer from $\AV\sim1.2$ mag
extinction in the nuclear region. It is also well known that K band
light is a better tracer of mass.  Therefore it is useful to use both
the HST and ground based light profiles in order to infer the mass
profiles.  The star cluster is probably washed out by the lower
spatial resolution in the NICS image, thus, in order to account for
its presence one can determine its geometrical parameters from the fit
of the HST light profiles and use them in the fitting the NICS image.

In principle, the WFPC2 images could be reddening corrected by using the
color maps in Figure \ref{fig:colours} to derive the $E(B-V)$ map. In order to
do this, one might assume that the colors in the nuclear region are constant
and equal to those beyond 10\arcsec\ (see Figure \ref{fig:profiles}). For a
more detailed description of this procedure see for instance
\citealt{marconi00}.  However the cluster would be forced to have the same
intrinsic
color as the galactic disk, which is probably not realistic. Moreover its shape
would be affected by PSF differences among the images.  Finally it is not clear
if the color differences are due entirely to reddening.  We therefore decided
not to apply any reddening correction and the consequences of this will be discussed in \S\ref{sec:discussion}.

We first consider the light profile derived from
WFPC2/F814W because it is the least sensitive to reddening among the
HST images.
We then fit a model light distribution composed of the
central star cluster and the more extended component. We assume
that the star cluster luminosity density is spherically symmetric with
a density law:
\begin{equation}
\rho^\star (r) = \rho^\star_b
\left( 1+\left(\frac{r}{r^\star_b}\right)^2\right)^{-\beta^\star}
\end{equation}
The extended component has a functional form of the type
\begin{equation}\label{eq:dens}
\rho (m) = \rho_b\left(\frac{m}{m_b}\right)^{-\alpha}
\left( 1+\left(\frac{m}{m_b}\right)^2\right)^{-\beta}
\end{equation}
where $m^2 = x^2+y^2+z^2/q^2$ corresponds to the radius in the
spherical case and $q$ is the axial ratio of the mass distribution.
This functional form of the extended component is a reasonable
description of the central part of the galaxy where the bulge
dominates. It is sufficient for our purposes because we are
interested in modeling only the inner $r<5\arcsec$.

\begin{figure*}[t!]
\centerline{
 \epsfig{figure=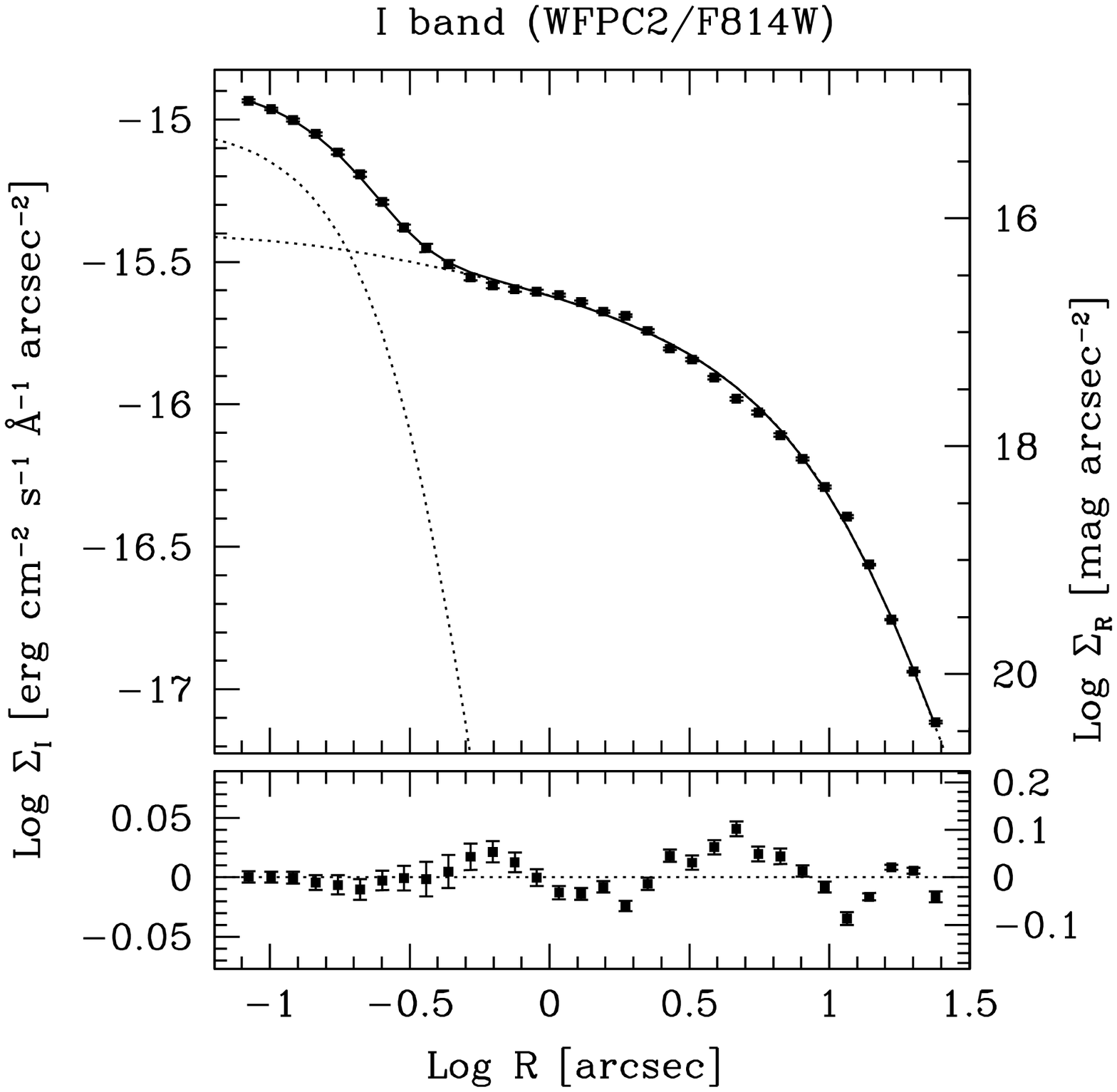,angle=0,width=0.45\linewidth}
 \hskip 0.05\linewidth
 \epsfig{figure=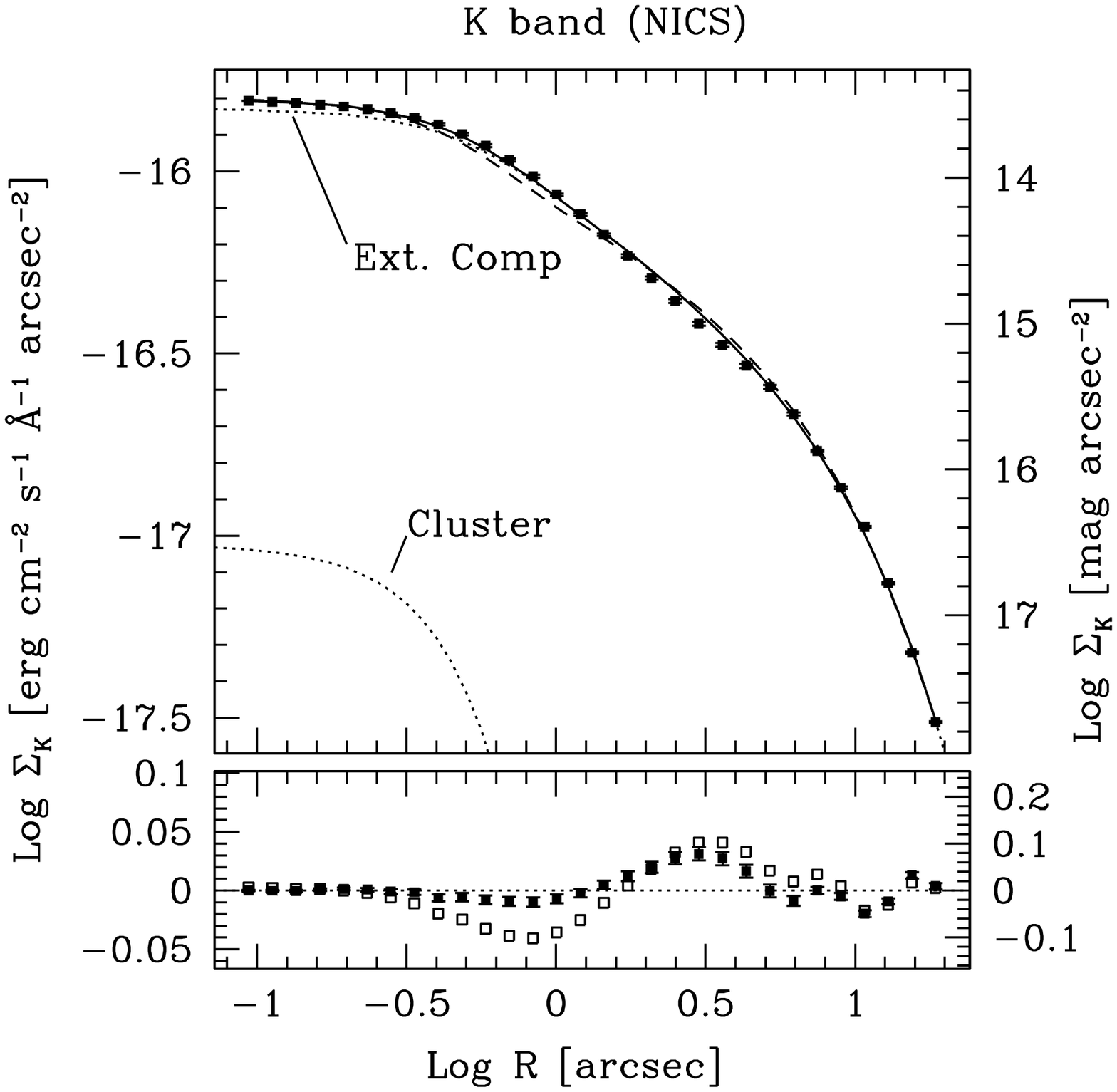,angle=0,width=0.45\linewidth}
}
\caption{\label{fig:fitprofs} Left: fit of the light profile obtained from
the WFPC2/F814W data.  Right: fit of the
NICS/K light profile with the central star cluster and the extended component
(the geometrical parameters of the star cluster were determined in the previous
fit). In both cases the lower panel shows the fit residuals. The dashed line
in the right panel shows
the K band fit obtained by fixing the cluster normalization to a value which is 0.6 dex larger than the best fit value. The empty squares
show the corresponding residuals.}
\end{figure*}

In Equation~\ref{eq:dens}, $q=1$ for a spherical light distribution, and
$q=0.1$ for a disk-like one. These are the two extreme cases.
The best fit parameters for the HST light profile are shown 
in the upper part of Table~\ref{tab:fitprofs}.  
The left panel in Figure~\ref{fig:fitprofs} 
shows the fit in the spherical case which is visually indistinguishable
from the disk case.

Using the geometrical parameters of the cluster determined from the 
fit to the HST data
($r^\star_b$ and $\beta^\star$) we fit the NICS profile, both the
spherical and disk case.  The results are shown in Table~\ref{tab:fitprofs}. The
right panels in Figure~\ref{fig:fitprofs} show the fit in the spherical case
which, as before, is visually indistinguishable from the disk-like case.

The values of the reduced \chisq\ reported in Table~\ref{tab:fitprofs}
are much larger than the expected value of 1 and the main reason can
be found in the systematic deviation of the residual around 3\arcsec\
(Figure~\ref{fig:fitprofs}) where the data are systematically
lower than the model. This systematic deviation, whose maximum value
is $\sim 0.03$ dex, i.e.\ 7\%, is present in both the optical and
near-IR light profile and is caused by the presence of the bar-like
feature.

\begin{figure}[t!]
\epsfig{figure=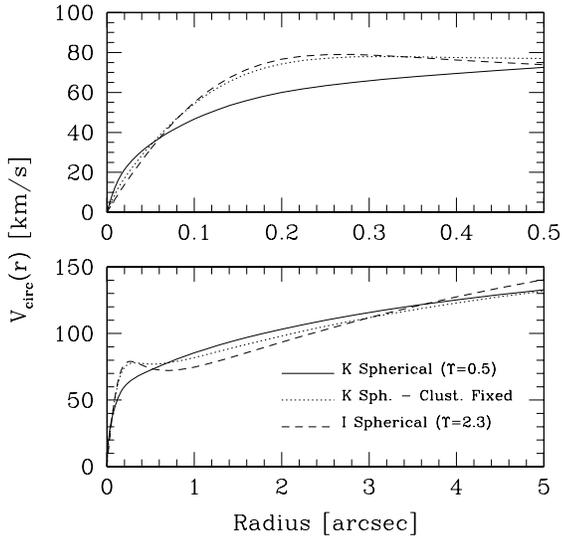,angle=0,width=\linewidth}
\caption{\label{fig:vels} Circular velocities in the principal plane of the potential for the density distributions determined from the fit of the NICS
and HST data.
The circular velocities are derived assuming a spherically symmetric
mass distribution and are plotted for $\mlr_K=0.5$
and $\mlr_I=2.3$. The dotted line represent the extreme case in which the
cluster normalization in the K band has been fixed to $+0.6$ dex of the
best fit value (see text).} 
\end{figure}

The fits to the light profiles allow an estimate of the I and K magnitudes of
the star cluster. With the assumed distance, the cluster luminosity is
$\ten{7.3}\,{\Lsun}_{wI}$ where $wI$ refers to the F814W band (in this band the
Sun has ${\Msun}_{wI}=4.16$). In the K band, the cluster has $\ten{7.2}\,
{\Lsun}_{K}$ for the spherical and $\ten{7.1}\,{\Lsun}_{K}$ for the disk cases
(${\Msun}_{K}=3.286$).
These results suggest that the cluster could be blue, with estimated
color $I-K\simeq 0.9$ (or 0.5 in the disk case).
The estimate of the K mag of the cluster is however uncertain due to the
fact that the cluster is unresolved in the ground based observations.
In order to set a limit to the amount of the cluster contribution to the 
K band light profile, in Figure~\ref{fig:fitprofs}
we also plot the results of the fit when the
cluster normalization has been fixed to a value 0.6 dex larger (i.e.\ the
cluster is 1.5 magnitudes brighter) than the best fit value (dashed line). 
The fit is worse, as clearly indicated by the residuals,
and we consider this a firm upper limit for
the cluster contribution to the K band light profile.
This implies that the observed color of the cluster is $I-K< 2.4$.
If we assume that the cluster is subjected to $\AV=1.2$ mag 
of extinction (\S\ref{sec:profiles}),
the upper limit changes to $I-K< 2.0$.

Figure~\ref{fig:vels} shows the circular velocities in the principal plane of the
potential expected from the density distributions determined from the fit of
the NICS and HST data. In order to match the different rotation curves the
circular velocities are plotted for $\mlr_K=0.5$ or $\mlr_I=2.3$, the values
derived in \S\ref{sec:modelkin} from fitting the rotation curves. It is clear
that the rotation curves are very similar beyond $r>0.5\arcsec$. However they
differ at smaller radii because of
the different contribution of the star cluster to
the total mass budget. This difference is due to the fact that the
mass-to-light ratio has been assumed constant in each band.

\subsection{\label{sec:modelkin}Kinematics}

In order to model the gas kinematical data (velocities and widths measured
along the slit) we select the simplest possible approach and we assume that
the ionized gas is circularly rotating in a thin disk located in the principal
plane of the galaxy potential. The latter assumption is not needed if the
galaxy potential has a spherical symmetry.  We assume that the disk is not
pressure supported and we neglect all hydrodynamical effects. Thus the disk
motion is completely determined by the gravitational potential which is made of
two components: one is stellar and is completely determined by the mass
distribution, derived in the previous section, and by its mass-to-light ratio
\mlr.  The other one comes from a dark mass concentration (the black hole),
spatially unresolved at HST+STIS resolution and defined by its total mass \MBH.
This is the same approach which has been followed in all previous gas
kinematical studies of circumnuclear gas disks  
\citep[\eg][]{macchetto97,marel98,barth01,marconi01}.
\begin{figure}[t!]
\epsfig{figure=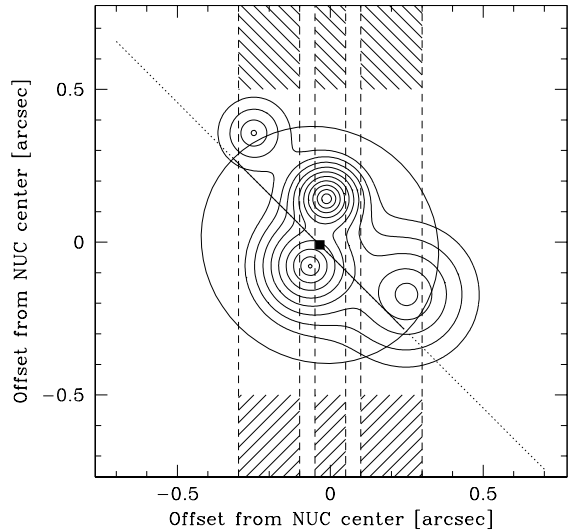,angle=0,width=\linewidth}
\caption{\label{fig:imflux} Contours of the modelled emission line flux
distribution of the nuclear gas disk with overplotted the slit
positions (vertical dashed lines), the position of the kinematic
center (filled square), and the line of nodes derived from fitting the
kinematic data (solid straight line).  The dotted line is the line of
nodes of the extended material. The ellipse indicates the nuclear
gas disk outer limit for an inclination of 20\arcdeg\ with respect to the
line of sight.}
\end{figure}
\begin{figure}[t!]
\epsfig{figure=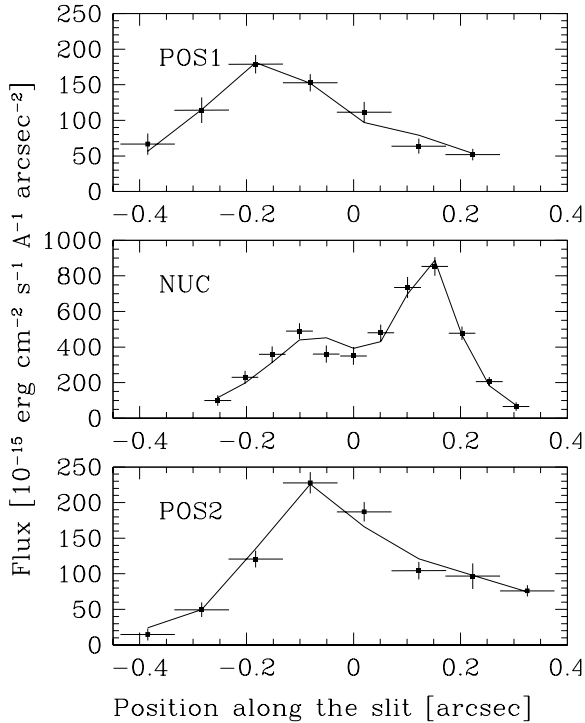,angle=0,width=\linewidth}
\caption{\label{fig:fluxnuc} The assumed flux distribution compared with the
data after folding with the telescope and instrument responses.
The model values are connected by straight lines in order to guide the eye.
} 
\end{figure}

In principle, in order to constrain the \BH\ mass, one should compare the
emission line profile predicted by the model with the observed one.  Even in
the simple potential described above, the line profiles can be very
complicated, with multiple peaks \citep{witold01}. 
Nevertheless, as shown in Figure~\ref{fig:fits}, the line profiles are well
represented by single gaussians (after subtracting the ``blue'' component)
and therefore we compare 
the moments of the line
profiles: the average velocity $\vel$ and velocity dispersion $\sigma$ defined
as $\sigma^2=\velsq-\vel^2$.  In order to be compared with the observations, the
model $\vel$ and $\sigma$ are computed taking into account the spatial
resolution of HST/STIS and the size of the apertures over which the observed
quantities are measured.  The formulae used to compute velocities, widths and
line fluxes and the details of their derivation, numerical computation and
model fitting are described in Appendix~\ref{app:gaskin}.  To derive the
observed $\vel$ and $\sigma$, our approach is to parameterize the observed
spectra by fitting them with as many gaussian components as
required by the data.  Then, components which are not obviously coming from
circularly rotating gas can be discarded \citep[\eg][]{winge99} and the
average velocities and line widths can be computed from the remaining ones
(this is trivial if, as in the present case, only a single gaussian component
is left).  With this approach, the  observed and model parameters, are
compared.  The advantages of our approach are twofold. Firstly, much
computational time is saved in computing just $\vel$ and $\sigma$ instead of the
whole line profile.  Secondly, modeling the observed emission line profiles in
detail requires very high S/N data and it is seldom possible to obtain such
high quality data with HST except for a few galaxies \citep[\eg\ M87,][]{macchetto97}.  Even with the high S/N available from 8m class telescopes the matching
of the line profiles remains a significant problem 
\citep[\eg\ Cen A,][]{marconi01}.

Thus, the model $\vel$ and $\sigma$ depend on $\Sigma$, the intrinsic surface brightness distribution of the emission lines, and on the following
parameters:
\begin{itemize}
\item \So, the position of the kinematical center along the slit with respect
to the position of the continuum peak;
\item \B, the distance between the reference slit center of the NUC slit and
the kinematical center;
\item \I, the inclination of the rotating disk;
\item \Th, the angle between the disk line of nodes and the slits;
\item \Vsys, the systemic velocity of the disk;
\item \mlr, the mass-to-light ratio of the stellar population;
\item \MBH, the \BH\ mass. 
\end{itemize}
Not all of these parameters can be independently determined by fitting the observations.

\begin{figure*}[t!]
{\centering
\epsfig{figure=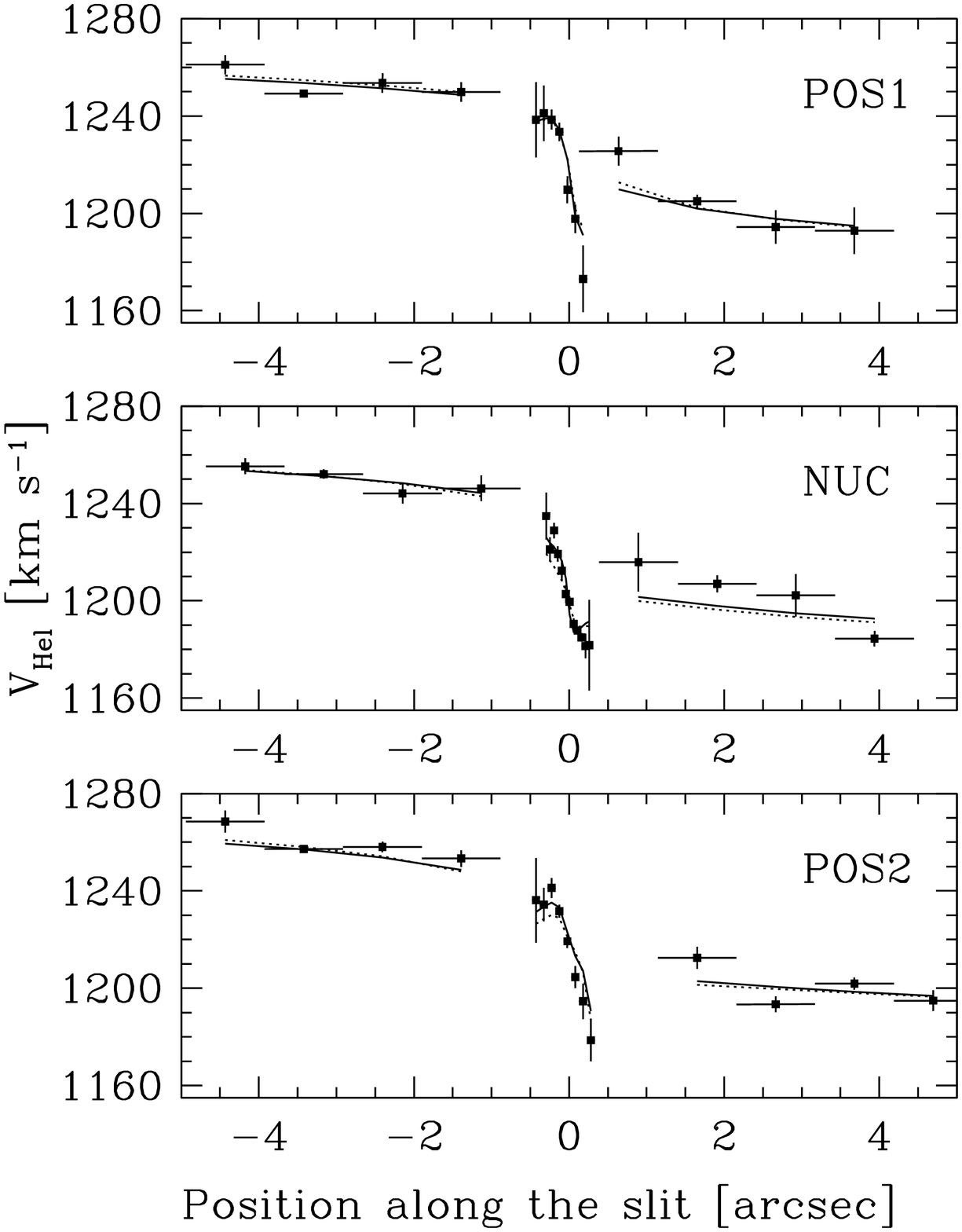,angle=0,width=0.49\linewidth}
\epsfig{figure=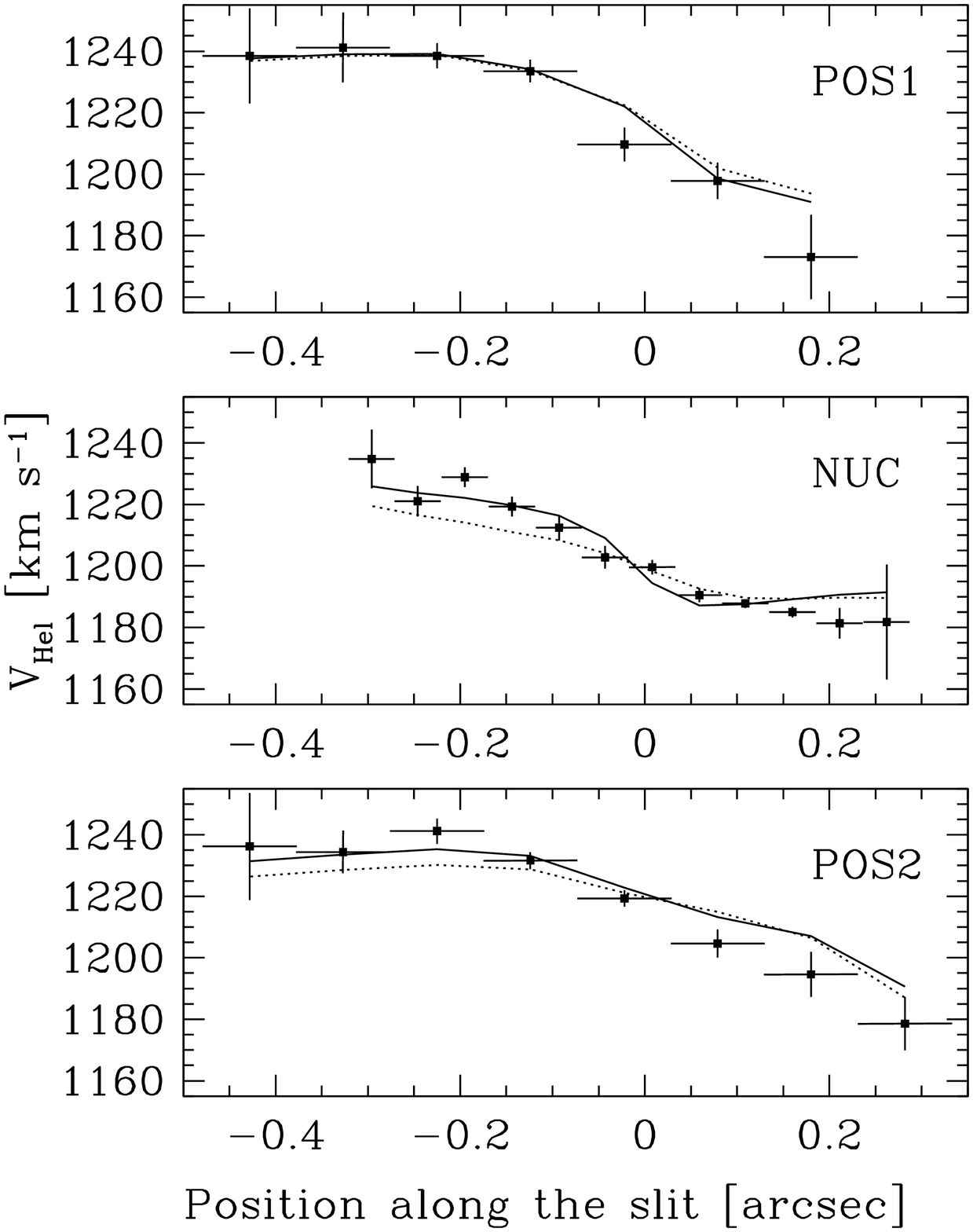,angle=0,width=0.49\linewidth}
}
\caption{\label{fig:fitstd} Best fit standard model of the observed rotation
curves (solid line) compared with the data. The dotted line is the best fit
model without a black hole.
The model values are connected by straight lines in order to guide the eye.
Note that points from external and nuclear regions are not
connected because they are kinematically decoupled.}
The right panel is a zoom on the nuclear disk
region. The plotted model uses the mass density distribution derived from
the K band light profile with the assumption of spherical symmetry. 
\end{figure*}

It can be inferred from the equations in Appendix~\ref{app:gaskin}
that \MBH, \mlr\ and \I\ are
directly coupled. This fact has already been discussed in detail by
\citet{macchetto97} but here we briefly indicate how this coupling can be used
to pose a
lower limit on the disk inclination \I.
Considering the case where no \BH\ is present, the amplitude of the rotation
curves depends linearly on $\mlr^{0.5}\sin\I$ and its value can be determined
by fitting the observations. Therefore \mlr\ will increase with decreasing \I. 
An upper limit on \mlr\ can be certainly set by the largest reasonable
value for a very old stellar population \citep[\eg][]{bc96,maraston}
and this will immediately fix the lower limit for \I. If a \BH\ is present,
$\mlr^{0.5}\sin\I$ can still be determined from the kinematical data obtained
beyond the \BH\ sphere of influence. Of course, the underlying assumption
is that both \mlr\ and \I\ are constant.

\subsubsection{The intrinsic surface brightness distribution of emission lines}

The intrinsic surface brightness distribution of emission lines
$\Sigma$ is an important ingredient in the model computations because
it is the weight in the averaging of the observed quantities.  The
ideal situation would be to have an emission line image with spatial
resolution higher than the spectra. Not having that, \citet{barth01}
have used a deconvolved HST image to model their HST data. This helped
them to match the microstructure of the velocity field at large
radii, but it is important to remember that the information about the
BH mass comes predominantly from within the central unresolved source
of unknown luminosity distribution.  Our approach is to attempt to
match the emission line fluxes as observed in the spectra (or in
emission line images) together with a variety of synthetic
realizations of this unknown central emissivity distribution.  We
use this to estimate its impact on the derived value of \MBH.  Thus,
we assume simple analytical forms of the line flux distribution which
well match the observed one along the slit, after folding with the
telescope and instrument responses as described above.

It is clear from the right panel in Figure~\ref{fig:rotnuc} that the
flux distribution observed at the three slit positions is very complex
and cannot be described by a single radially symmetric component in
the disk plane.  Therefore we choose the approach of modeling directly
the flux distribution on the plane of the sky by exploring various
simple analytical forms of the line flux distribution.
We take the intrinsic light distribution at point $x,y$ in the plane
of the sky in the NUC slit reference frame ($x$ is the position across
the NUC slit while $y$ is the position along it) to be defined as
\begin{equation} I(x,y) = \sum_i I_{\circ i}\,
		f_i\left(\frac{r_i}{r_{\circ i}} \right)
\end{equation}
where $f_i$ is a circularly symmetric function of characteristic radius
$r_{\circ i}$ and weighting factor $I_{\circ i}$.  Each component function
$f_i$ is centered at $(x_{\circ i},y_{\circ i})$, and $r_i$ is the  radial
distance from this location $r_i = [(x-x_{\circ i})^2+(y-y_{\circ i})^2]^{0.5}$.
This is an approach similar to that of \citet{barth01} except that we use a
synthetic realization of the line flux map.
Although each bright patch is assumed to be circularly symmetric, its material
is in circular rotation about the galactic nucleus. Hence patches are
constantly shearing rather than moving as coherent units.

The free parameters characterizing the flux distribution are therefore
$(x_{\circ i}$, $y_{\circ i})$, $I_{\circ i}$ and  $r_{\circ i}$ and are chosen
in order to match the observed flux distribution along the slit after folding
the model with telescope and instrument.  For example, one of the functional
forms used 
is the combination of 4 exponentials and a constant baseline i.e.\ 
$f_i(r_i/r_{\circ i})=\exp(r_i/r_{\circ i})$. Figure~\ref{fig:imflux} shows the
contours of this line flux distribution which matches the observed profile along
the slit after convolving with the instrumental response (see Figure~\ref{fig:fluxnuc}).

\subsubsection{\label{sec:modstand}The standard approach}
\begin{deluxetable}{cccccccccc}
\tablewidth{0pt}
\tablecolumns{10}
\tablecaption{\label{tab:fitstandard} 
Model fit results in the standard approach.}
\tablehead{
 \colhead{Band} &
 \colhead{$q$} &
 \colhead{$\log \MBH$}&
 \colhead{\mlr}&
 \colhead{\So} &
 \colhead{\B} &  
 \colhead{\Th} &
 \colhead{\Vsys} &
 \colhead{$\Delta\Vsys$} &
 \colhead{\chisqr} \\
 \colhead{\small (a)} &
 \colhead{\small (b)} &
 \colhead{\small (c)}&
 \colhead{\small (d)}&
 \colhead{\small (e)} &
 \colhead{\small (e)} &  
 \colhead{\small (f)} &
 \colhead{\small (g)} &
 \colhead{\small (g)} &
 \colhead{\small (h)} } 
\startdata
K & 1.0 & 7.04    & 0.52 & $-0.04$ & 0.02 & $-45$ & 1212 & 11 & 2.32 \\
K & 1.0 & 0.0\tst & 1.00 & $-0.09$ & 0.06 & $-58$ & 1211 & 11 & 2.88 \\
K & 0.1 & 7.09    & 0.22 & $-0.07$ & 0.08 & $-32$ & 1213 & 11 & 2.13 \\
I & 1.0 & 6.86    & 2.28 & $-0.01$ & 0.00 & $-42$ & 1210 & 12 & 1.86 \\
I & 1.0 & 0.0\tst & 2.65 & $-0.03$ & 0.03 & $-45$ & 1210 & 12 & 2.09 \\
I & 0.1 & 7.14    & 1.65 & $-0.03$ & 0.01 & $-47$ & 1212 & 10 & 2.22 \\
\enddata
\tablenotetext{a~}{Band from which the mass density profile was derived.} 
\tablenotetext{b~}{Assumed axial ratio of the mass distribution.}
\tablenotetext{c~}{Log of \BH\ mass in units of \Msun.}
\tablenotetext{d~}{Mass-to-light ratio in used band.}
\tablenotetext{e~}{Arcsec}
\tablenotetext{f~}{Degrees}
\tablenotetext{g~}{\kms}
\tablenotetext{h~}{Reduced $\chi^2$.}
\tablenotetext{\star~}{The parameter was fixed to this value.}
\end{deluxetable}
\begin{figure*}[t!]
{\centering
\epsfig{figure=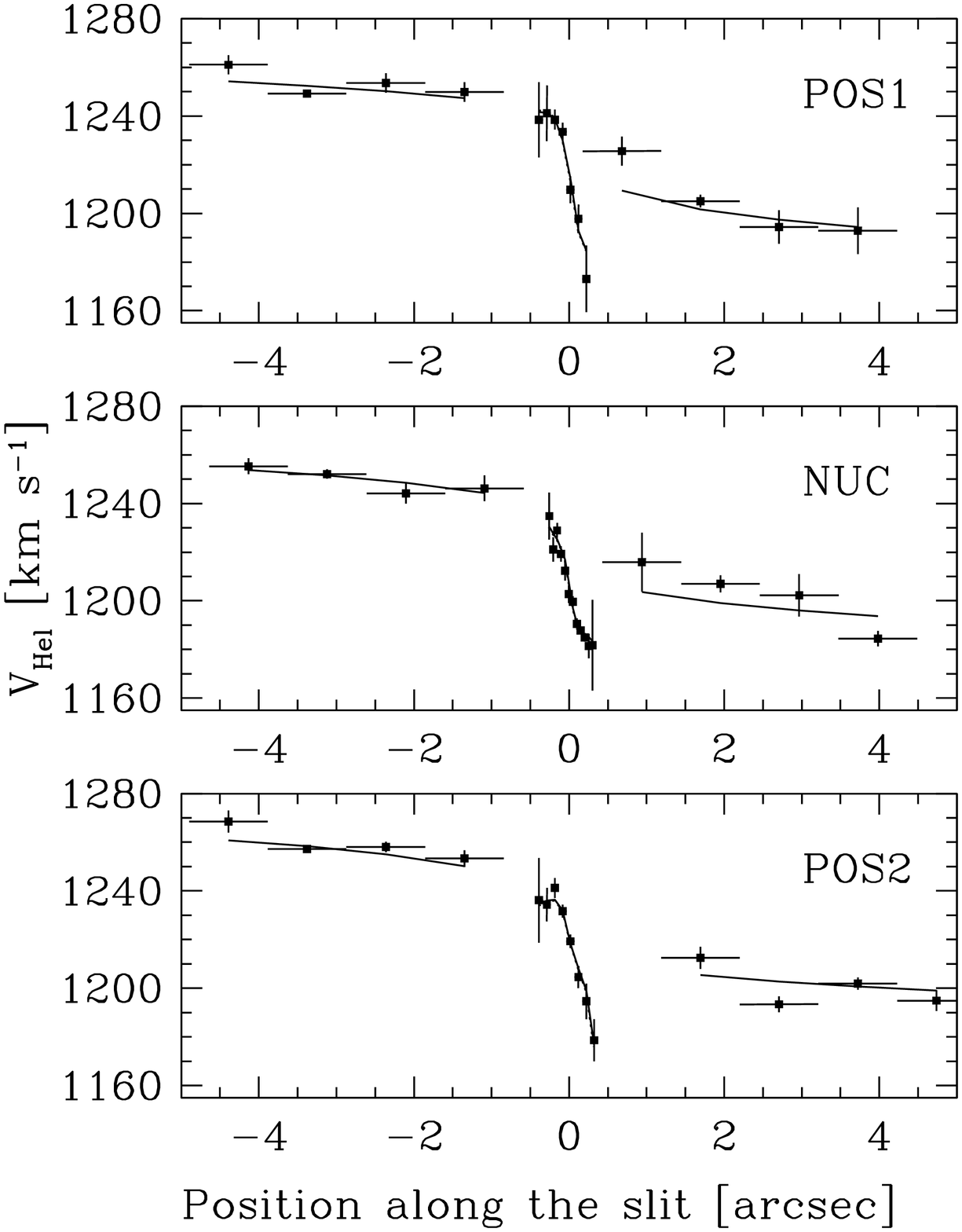,angle=0,width=0.49\linewidth}
\epsfig{figure=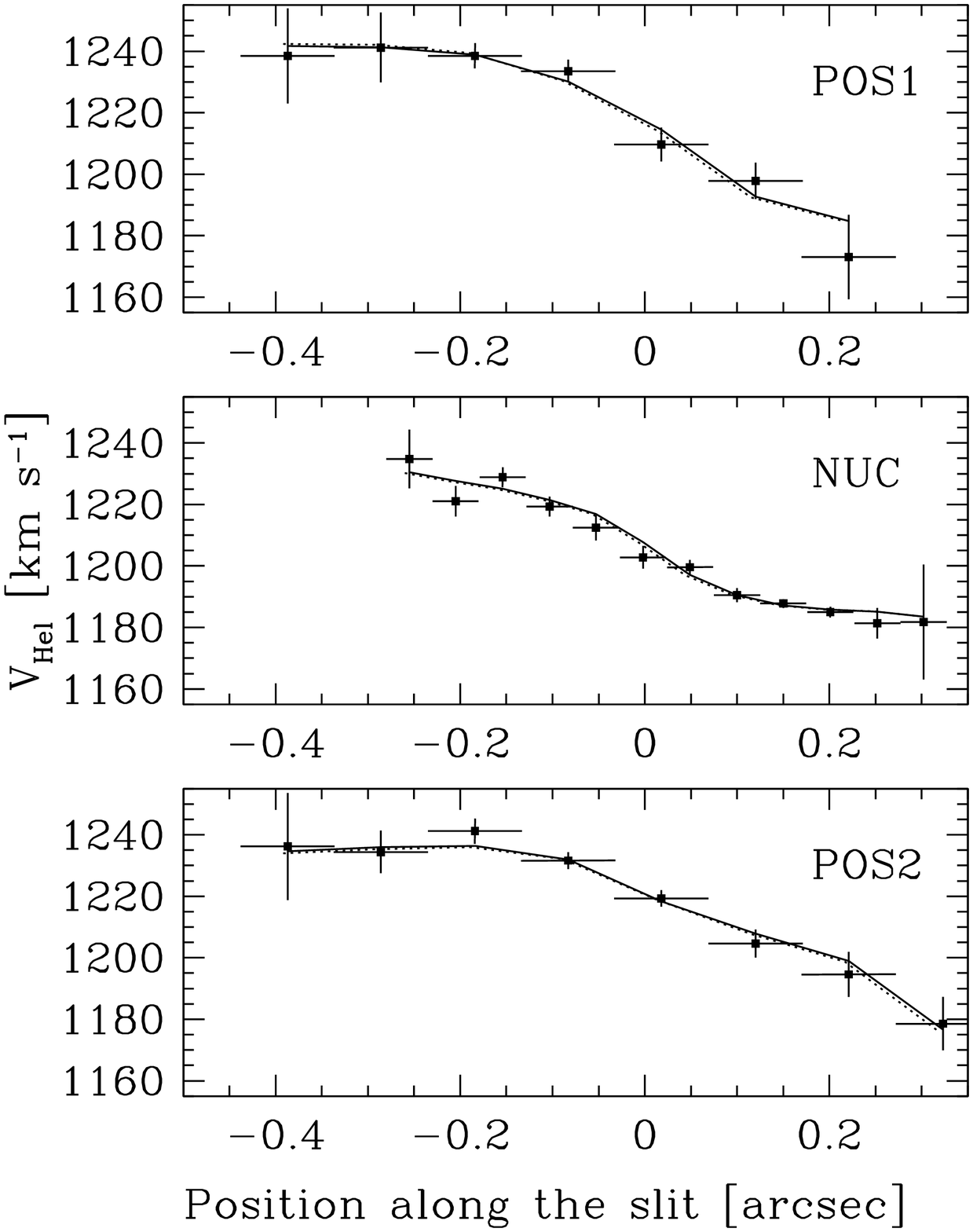,angle=0,width=0.49\linewidth}
}
\caption{\label{fig:fitaltern} Best fit alternative model of the observed
rotation curves compared with the data. The solid line is the best fit model
obtained by fixing \I\ and varying \mlr\ while the dotted line represent the
case in which \I\ has been varied and \mlr\ kept fixed.
The model values are connected by straight lines in order to guide the eye.
Note that points from external and nuclear regions are not
connected because they are kinematically decoupled.
The right panel is a
zoom on the nuclear disk region. The plotted model uses the mass density
distribution derived from the K band light profile with the assumption of
spherical symmetry.} 
\end{figure*}

\begin{deluxetable}{cccccccccc}
\tablewidth{0pt}
\tablecolumns{10}
\tablecaption{\label{tab:fitaltern} 
Model fit results in the alternative approach.}
\tablehead{
 \colhead{Band} & \colhead{$q$} & \colhead{$\log \MBH$}& \colhead{\mlr}&
 \colhead{\So} & \colhead{\B} &  \colhead{\Th} & \colhead{\I} &
 \colhead{\Vsys} & \colhead{\chisqr} \\
\colhead{\small (a)} & \colhead{\small (b)} & \colhead{\small (c)}& \colhead{\small (d)}& \colhead{(e)} &
 \colhead{\small (e)} &  \colhead{\small (f)} & \colhead{\small (f)} & \colhead{\small (g)} & \colhead{\small (h)} } 
\startdata
\cutinhead{Extended Component}
K & 1.0 & 0.0\tst & 0.48  &0.0\tst &0.0\tst & $-43$\tst & 20\tst & 1224 & 2.7 \\
K & 0.1 & 0.0\tst & 0.31  &0.0\tst &0.0\tst & $-43$\tst & 20\tst & 1224 & 2.7 \\
I & 1.0 & 0.0\tst & 2.29  &0.0\tst &0.0\tst & $-43$\tst & 20\tst & 1224 & 2.5 \\
I & 0.1 & 0.0\tst & 1.38  &0.0\tst &0.0\tst & $-43$\tst & 20\tst & 1224 & 2.5 \\
\cutinhead{Nuclear Disk}
K& 1.0& $<$6.8& 1.29    & 0.0     & $-0.05$   & $-40$  & 20\tst & 1204 & 0.66 \\
K& 1.0& $<$6.6& 0.48\tst& $-0.01$ & $-0.04$   & $-39$  & 35     & 1205 & 0.67 \\
K& 0.1& $<$6.8& 0.88    & $-0.01$ & $-0.05$   & $-40$  & 20\tst & 1205 & 0.69 \\
I& 1.0& $<$6.8& 4.05    & $-0.01$ & $-0.04$   & $-41$  & 20\tst & 1206 & 0.60 \\
I& 0.1& $<$6.8& 3.77    & 0.00    & $-0.04$   & $-41$  & 20\tst & 1205 & 0.58 \\
\enddata
\tablenotetext{a~}{Band from which the mass density profile was derived.} 
\tablenotetext{b~}{Assumed axial ratio of the mass distribution.}
\tablenotetext{c~}{Log of \BH\ mass in units of \Msun.}
\tablenotetext{d~}{Mass-to-light ratio in used band.}
\tablenotetext{e~}{Arcsec}
\tablenotetext{f~}{Degrees}
\tablenotetext{g~}{\kms}
\tablenotetext{h~}{Reduced $\chi^2$.}
\tablenotetext{\star~}{The parameter was fixed to this value.}
\end{deluxetable}

\begin{figure*}[t!]
{\centering
\epsfig{figure=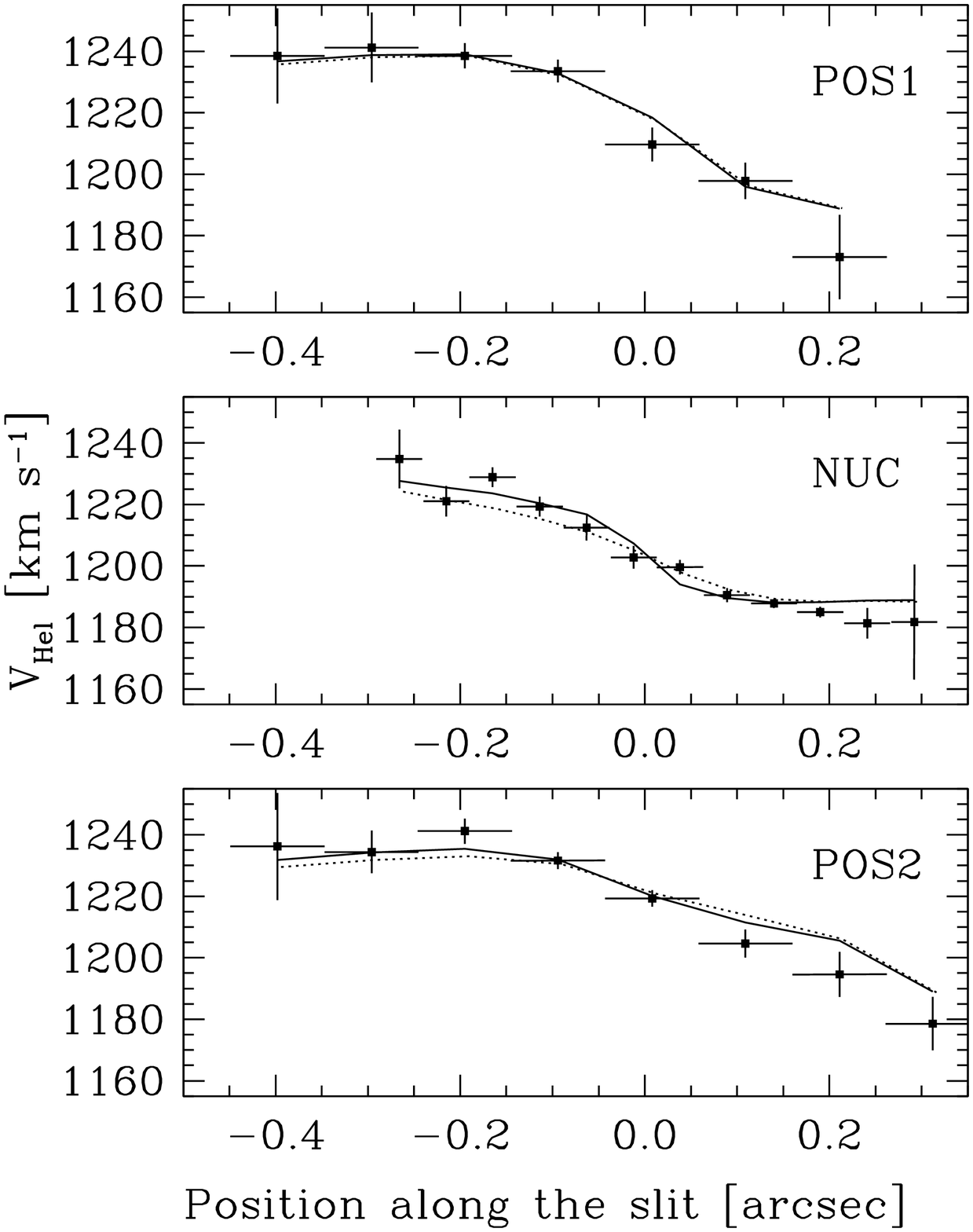,angle=0,width=0.49\linewidth}
\epsfig{figure=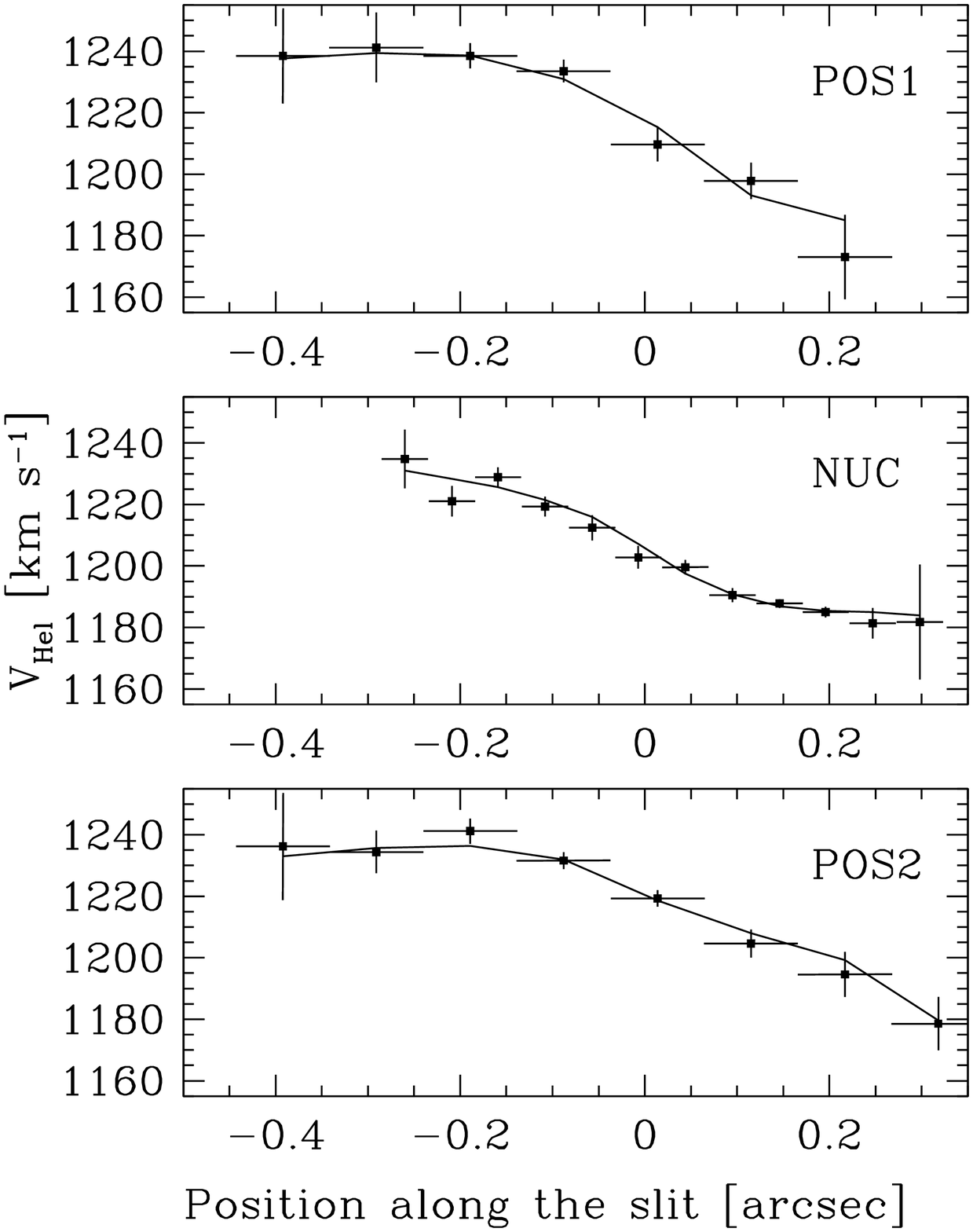,angle=0,width=0.49\linewidth}
}
\caption{\label{fig:fitI} Best fit models of the observed rotation
curves computed using the mass densities derived from the I band light profiles.
The model values are connected by straight lines in order to guide the eye.
The left panel refers to the standard model (the dotted line is the model
without a \BH) while the right panel refers to the alternative model.} 
\end{figure*}

The standard approach followed so far in gas kinematical analysis is to assume
that (i) gas disks around black holes are not warped i.e.\ they have the same
line of nodes and inclinations as the more extended components, and (ii) the
stellar population has a constant mass-to-light ratio with radius
\citep[\eg][]{marel98,barth01}.  In the present case, the blueshift of the
inner disk (\S\ref{sec:vcurves}) indicates that the standard approach must
generalized to allow for the kinematical decoupling between inner and large
scale disks.

We use the emission line flux distribution derived in
the previous section and we fix the inclination to \I=20\arcdeg, i.e.\ the
inclination of the galactic disk.  The free parameters of the fit are
then \MBH, \mlr, \So, \B, \Th, \Vsys\ and $\Delta\Vsys$, the velocity
shift of the extended component with respect to the nuclear one.  We
perform the fit using the mass density profiles derived for the I and
K band, both in the spherical and disk cases.  The results of the fit
are shown in Table~\ref{tab:fitstandard}.  Statistical errors at the
$2\sigma$ level on $\log\MBH$ and $\log\mlr$ are $\pm 0.2$ and $\pm
0.1$, respectively, thus the derived \BH\ mass is
$\MBH=1^{+0.6}_{-0.7}\xten{7}\Msun$.  Figure~\ref{fig:fitstd} shows the
best fit model (solid line) obtained with the mass density
distribution derived from the K band light profile with the assumption
of spherical symmetry. The dotted line is the best fit model without a
black hole.  The left panel of Figure~\ref{fig:fitI} shows the fit of
the NUC data from the model with the mass distribution derived from
the I band.  Figures~\ref{fig:fitstd} and \ref{fig:fitI}, indicate that
a model with no \BH\ cannot account for the observed nuclear rotation
curve, producing a velocity gradient which is shallower than observed.
Note that the position angle of the line of nodes, a free parameter of
the fit, is the same as the one inferred from the large scale CO
velocity map by \citet{sakamoto99}. This supports the assumption that
the nuclear disk is the continuation at small scales of the
galactic disk.  However the fit confirms that the nuclear disk is
blueshifted by $\sim 10\kms$ with respect to the extended component
and this argues against the assumption in previous sentence.  
The fit is greatly improved if a velocity shift of 8\kms\ is allowed for
the POS2 data. This velocity shift is the consequence of an error on 
the absolute wavelength calibration of the POS2 data and is well
within the expected STIS performances \citep{stishand}.

The value of \mlr\ in the K band derived from the fit varies between 0.2
(disk light distribution) and 0.5 (spherical light distribution).  This
range of values is in good agreement with the typical K-band mass-to-light
ratios of spiral bulges \citep{moriondo98}. 
Similarly the value of \mlr\ in the I band ranges between 2.2 and 3.6
in agreement with measurements within the inner 2\KPC\ of spiral
galaxies \citep{giovanelli02}.

\subsubsection{\label{sec:modalt}Alternative model: the decoupled nuclear disk}

To date, everyone who has determined a central \BH\ mass from gas kinematics
assumed that the disks are unwarped, i.e.\ coplanar at all radii,
and that the stellar mass-to-light ratio is constant with radius.

The high surface brightness and velocity offset of the nuclear disk with
respect to the extended component presented in \S\ref{sec:results} indicate
that, at least for NGC 4041, this assumption could be untrue.  The nuclear and
large scale disks might be characterized by different geometrical properties
like position angle of the line of nodes, inclination and systemic velocity.
Recently \citet{cappellari02} have shown a discrepancy between the \BH\ mass in
IC 1459 determined from gas kinematical \citep{verdoes00} and stellar dynamical
models. The authors propose that a possible solution to this discrepancy could
be an error on the assumed gas disk inclination.

Therefore, in the alternative approach presented here, we first fit the extended
component data in order to determine the mass-to-light ratio \mlr\ of
the stellar component to be used in the fit of the nuclear data.  For
the extended component we assume a constant line flux distribution and
determine \mlr\ in the two extreme cases of the spherical and
disk-like light distribution (i.e.\ with an axial ratio of $q=0.1$).
\begin{figure}[t!]
\epsfig{figure=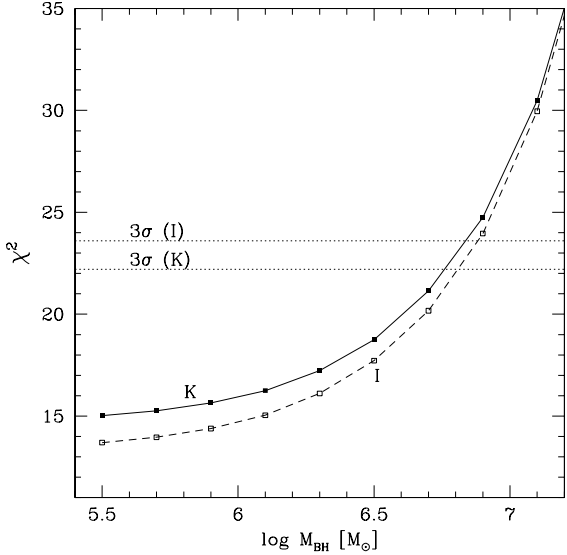,angle=0,width=\linewidth}
\caption{\label{fig:uplim} Statistical upper limits on the \BH\ mass in the
alternative approach for the cases in which the mass density has been derived
by the I and K band light profiles.} 
\end{figure}

From the CO velocity map by \citet{sakamoto99} the PA of the kinematic
line-of-nodes is PA$=86$\arcdeg, i.e.\ almost exactly East--West. Since the
PA of the slit in the STIS observations is 43\arcdeg, this means that the
angle between the slit and the disk line of nodes is
$\Th=-43\arcdeg$ (for the definition of \Th\ see Appendix~\ref{app:gaskin}). We
also assume that the inclination of the disk is \I=20\arcdeg, as specified
in the previous section.  The CO velocity map also clearly indicates
that the large scale velocity field is circularly symmetric.

The fitted circular rotation curves of the extended gas are shown in Figure~\ref{fig:fitaltern}. The derived \mlr\ for the spherical or disk geometry are
shown in Table~\ref{tab:fitaltern} for the mass density profiles
derived from both the K and I band light.  In our
derivation we allowed for a constant velocity shift between the data points in
the off-nuclear slit positions and the nuclear one. This is caused by the
uncertainty in the absolute wavelength calibration, and is  very small, -1\kms\
and 8\kms\ for POS1 and POS2 respectively. 
\begin{figure}[t!]
\epsfig{figure=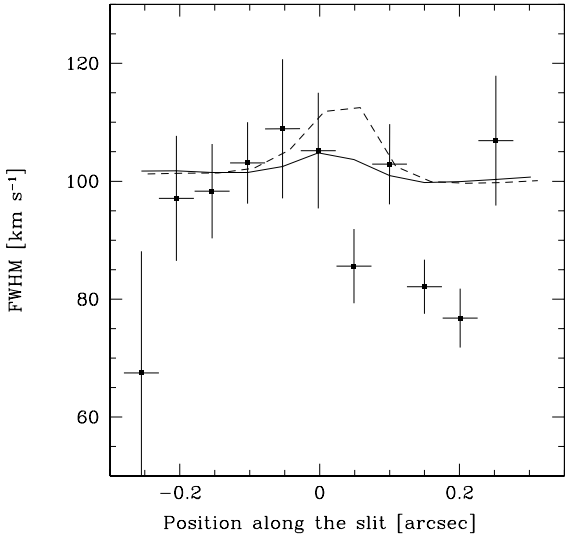,angle=0,width=\linewidth}
\caption{\label{fig:fwhmnuc} Observed and model FWHMs at the NUC
position. The solid line is the case without a BH (alternative model)
and the dashed line is with a \ten{7}\Msun\ \BH\ (standard model). The
assumed intrinsic velocity dispersion is 35\kms.}
\end{figure}

Similarly to the results of the standard model, mass-to-light ratios in the K
and I band are consistent with typical values for the central regions of spiral
galaxies.

We now fit the nuclear data ($r\le0\farcs4$) allowing for values of \mlr\
or \I\ different from those of the extended component.
Results of the fit are shown in Figure~\ref{fig:fitaltern} and
in Table~\ref{tab:fitaltern}. The right panel of Figure~\ref{fig:fitI}
shows the fit of the NUC data from the model with the mass
distribution derived from the I band.

In all these cases the \BH\ mass has been set to zero and the goodness
of the fit suggest that no \BH\ is needed to fit the data with
this approach.  We estimate the upper limit on the \BH\ mass by a 1
parameter variation, i.e.\ we assign a fixed value of \MBH\ and we
perform the fit.  When we have $\Delta\chi^2>=1,4,9$ we have reached
the 1,2,3 $\sigma$ upper limit.  As an example, in
Figure~\ref{fig:uplim}, we show the $\chi^2$ plot corresponding to the K
and I band spherical cases. There the 3$\sigma$ upper limit is
\ten{6.8}\Msun.

From Table~\ref{tab:fitaltern} it can be seen that the position of the
nucleus along and across the slit is independent of the assumed mass
distribution. The inclination varies, but like \mlr, is just a scaling
factor. Indeed varying \mlr\ produces the same results, as
shown in the same table.  The angle between the slit and the line of
nodes is very well determined in the range $-39 - \, -41$\arcdeg.
This means that
the PA of the kinematical line of nodes is 43\arcdeg\ (the PA of the
slit) minus $-40\arcdeg$ i.e.\ $\sim 83$\arcdeg\ consistent with the PA of the
kinematical line of nodes on large scales.  The systemic velocity is
$1204 - 1206\kms$ and is definitely blueshifted with respect to the extended
rotation where it is 1224\kms.

As pointed out by \citet{barth01} the FWHM of the lines might be a worrisome
issue in the sense that FWHM much larger than instrumental might indicate
motions which could deviate from circular (apart from the broadening of the
line due to unresolved rotation around a point mass).  In Figure~\ref{fig:fwhmnuc} we plot the observed FWHMs and compare them with the expected
one in the framework of the alternative model
assuming an intrinsic velocity dispersion of only 35\kms.
Also the dashed line represents
the case with a \ten{7}\Msun\ \BH\ showing that, in this case, the FWHM
cannot pose good constraints on the \BH\ mass.

\section{\label{sec:discussion}Discussion}
\begin{figure}[t!]
\epsfig{figure=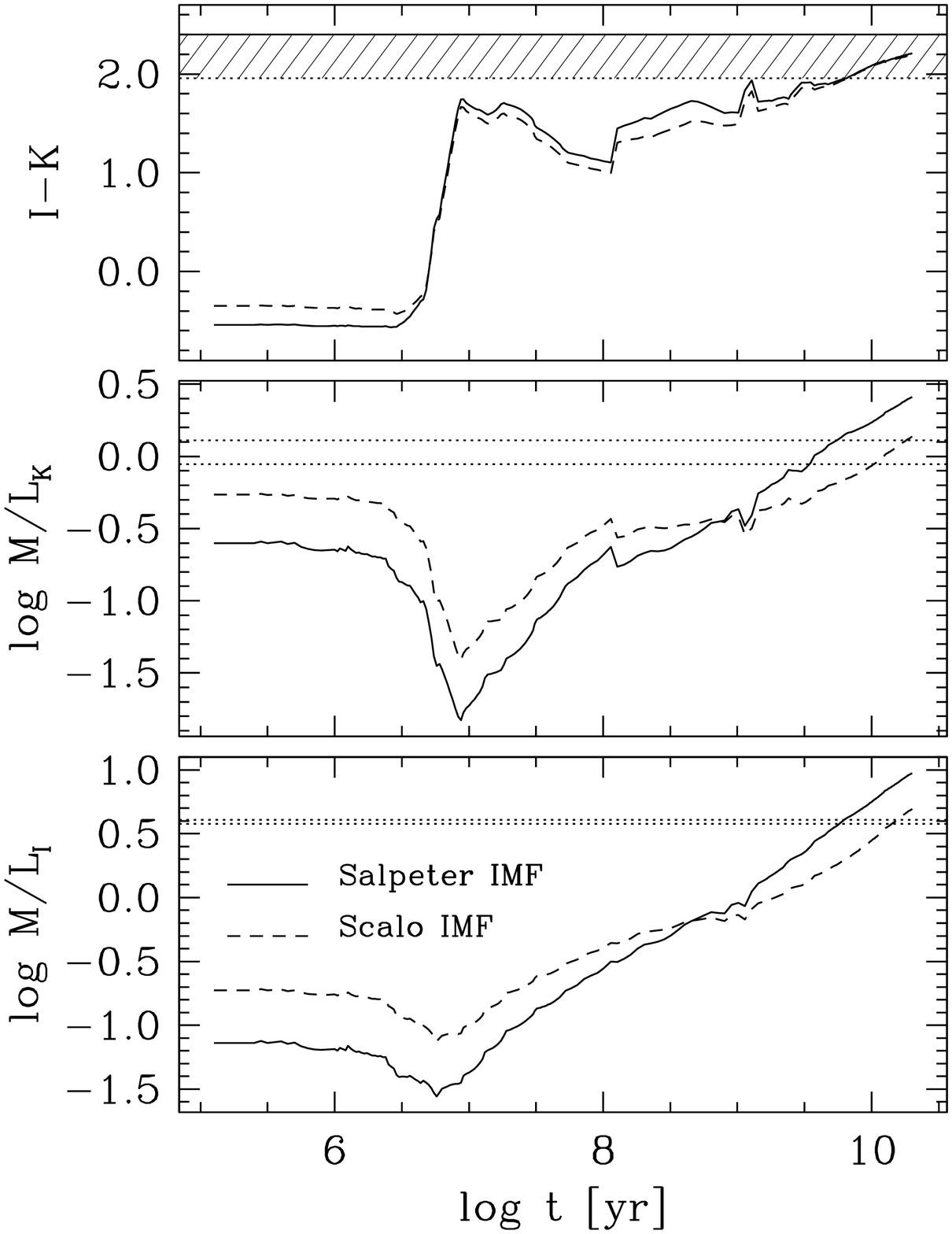,angle=0,width=\linewidth}
\caption{\label{fig:bc96} From top to bottom: $I-K$, $\log M/L_K$ and
$\log M/L_I$ vs time for a single stellar population (instantaneous
burst) with solar metallicity (models by \citealt{bc96}, updated in
1996).  The solid and dashed line are for a \citet{s55}
and a \citet{sc86} IMF
respectively. In the upper panel, the solid horizontal line represents the
observed upper limit on the $I-K$ color.  The dotted line is the same upper
limit after correction for $\AV=1.2$ mag. In the central and lower panel, the
horizontal dotted lines indicates the range of mass-to-light ratios allowed by
the observations. 
}
\end{figure}

\subsection{How the \BH\ mass determination is affected by assumptions.}

In \S\ref{sec:modelkin} we have shown that the inclination of the nuclear disk,
\I, and the mass-to-light ratio of the stellar population, \mlr, play a
critical role in detecting the presence of a nuclear \BH.  The {\it standard}
model, in which \I\ and \mlr\ are the same for the nuclear disk and the more
extended galactic disk, requires the presence of a \BH\ with $\MBH=1^{+0.6}_{-0.7}\xten{7}\Msun$
(\S\ref{sec:modstand}).  Conversely, allowing either \mlr\ or \I\ to vary, the
kinematical data can be fit without a \BH\ (the so-called {\it alternative}
model, \S\ref{sec:modalt})
and $\MBH<6\xten{6}\Msun$.  These results show a caveat in gas
kinematical searches, where it is always assumed
that \I\ and \mlr\ are constant at all radii.

To distinguish between the standard and alternative models it is necessary to
find some extra physical constraints. In particular the value of mass-to-light
ratio \mlr\ determined from the observations should be consistent with
realistic stellar populations
(see \S\ref{sec:modelkin}).  Little can be said about \I\ which in contrast
with the other geometrical parameter of the disk, \Th, is poorly
constrained by the rotation curves. Usually one can infer an estimate
of \I\ from the morphology of the disk observed in emission line
images,  especially in E and S0
galaxies \citep[\eg][]{marel98,barth01}. 
The root of the problem lies in the coupling between \MBH, \mlr\
and \I\ that arises because they are all essentially scaling factors
on the amplitude of the rotation curves.

In the {\it standard} approach (Table~\ref{tab:fitstandard}) \MBH\ does
not depend on the assumed mass density profile (either from the K or I
band).  The angle \Th, which is a free parameter of the fit, is quite
well constrained around $-40\arcdeg$, i.e.\ the PA of the kinematical line of
nodes is the same as the one at large scales derived from the 2D CO
map by \citet{sakamoto99}.  The constancy of \Th, which is not an
intrinsic parameter of the system but only depends on the galaxy
orientation in the plane of the sky, suggests that \I\ ($\Delta\I$
between the nuclear and galactic disk is an intrinsic parameter of the
system) should also be constant.  No similar argument can be placed on
\mlr\ which can then be allowed to vary and it is then possible to fit
the kinematical data without the presence of a \BH. This requires that
\mlr\ in the nuclear region is a factor of $\sim 2$ larger than that
in the extended disk.

To verify if this \mlr\ is consistent with the observed colors, we
plot in Figure~\ref{fig:bc96} $I-K$, $\mlr_K$ and $\mlr_I$ as a function
of time from the burst for a single stellar population experiencing an
instantaneous burst of star formation. We have used the models by
\citet{bc96}, updated in 1996, we considered solar metallicities,
\citet{s55} 
and \citet{sc86} IMF (solid and dashed lines in the figure), and
 used the theoretical stellar libraries.  The solid line in the
upper panel marks the upper limit we placed on the $I-K$ color of the
stellar cluster (\S\ref{sec:stardens}); the dotted line marks the same
upper limit but dereddened for $\AV=1.2$ mag (\S\ref{sec:profiles}).
Finally, the dotted lines in the central and lower panels shows the
\mlr\ required to fit the data in the alternative model both for the
spherical and disk-like distribution of stars. 
The \mlr s implied by the analysis are thus consistent with a very old
nuclear star cluster, with age $\sim 10$ Gyr.  This is apparently in contrast
with the high \HA\ equivalent width measured in the nuclear spectrum, $\simeq
70$ \AA, which seems to require the presence of very young stars.
A possibility is that the \HA\ emission is not due to young stars but to a low
luminosity AGN and, indeed, the total \HA\ luminosity in the nuclear region is
$L(\HA) = 4.9\xten{4}\Lsun$ corresponding to the low end of the $L(\HA)$
distribution for AGNs (\citealt{ho97}). To prove that the emission lines
are excited by an AGN, one could use the so-called diagnostic line ratios 
\NII/\HA\ and \OIII/\HB\ (\eg\ \citealt{osterbrock}). Ground
based spectra with 1\arcsec\ resolution (Axon et al.\ in preparation)
indicate that this object is on the
boundary between HII regions and LINERs with $\NII/\HA\sim 1$ and
$\OIII/\HB<0.5$. Unfortunately the spectra coverage of our 
HST data is limited to 6500-7000 \AA\ region but the
the $\NII/\HA$ ratio at 0\farcs1 resolution remains
similar to that observed from the ground suggesting that the same could 
be true for the \OIII/\HB.

If on the other hand
the star cluster is young, it can fully account for the \HA\ emission.
Assuming case B recombination, the 
$L(\HA)$ luminosity implies an ionizing photon rate of
$\QH\sim 2\xten{50} \SEC\1$ (\eg\ \citealt{osterbrock}),
which corresponds to $\sim 2 - 20$ O stars \citep{panagia73}.
When confronted with stellar population
synthesis model \citep{leitherer99} the observed \HA\ equivalent width indicates
an age of $\sim\ten{7}\YR$.
In this case, as shown in Figure \ref{fig:bc96}, the I band
mass-to-light ratio is more than a factor ten lower and the fit of the
kinematical data requires a \BH\ even in the alternative model. However
when using the stellar mass density profile derived from the K band, the data
can still be explained by stars alone. The reason is that the contribution of
the star cluster to the total light density profile is small, as can be seen
from the rotation curves in Figure \ref{fig:vels}.
Note that in order to
explain the high $\NII/\HA$ ratio with photoionization
from stars one needs to allow for higher-than-solar N abundances in the nucleus
as has been suggested many times in the past since the work by \cite{burbidge}.

In order to proceed further and firmly establish the age of the star cluster,
one needs HST/NICMOS infrared images of the nuclear region. 
The optical to near-IR colors of the star cluster constrain
its age and mass-to-light ratio. Moreover with HST/STIS blue spectra one can
measure the \OIII/\HB\ ratio, in order to estimate the AGN
contribution, and verify the existence of deep H Balmer lines in absorption
expected if  the cluster is young.

The other assumptions behind the modeling are the shape of the stellar density
profile (spherical vs disk-like, I vs K band, effects of reddening), the
intrinsic flux distribution of the emission lines and, finally, the assumption
of circular motions.

The stellar density profile depends on $q$, the intrinsic  axis
ratio of the extended mass distribution, and on the relative
importance between the nuclear star cluster and the extended stellar
component.  As shown in the previous section, the four cases examined
there cover the most extreme cases: spherical or disk-like
distribution of the extended component ($q=1$ and $q=0.1$
respectively), stellar cluster which gives a small contribution to the
total stellar density distribution (K band) or which dominates over
the nuclear disk size i.e.\  $r<0\farcs4$ (I band).  From
Tables~\ref{tab:fitstandard} and \ref{tab:fitaltern} we can conclude
that the value of the \BH\ mass or its upper limit do not depend on
the assumed stellar density profile.  From the same table it can be
seen that the position of the nucleus along and across the slit is
also independent of the assumed mass distribution.
In \S\ref{sec:stardens} we decided not to apply a reddening correction to
the light profiles because of the uncertainties involved.  This reddening
correction would have affected the values of the mass-to-light ratios derived
from the rotation curves (see \S\ref{sec:modelkin}) whose accurate
measurement is not the
aim of this paper.  Roughly, assuming an average $\AV=1.2$ mag
(\S\ref{sec:profiles}), the I light would increase by $\sim 70\%$ and $\mlr_I$
would decrease by the same amount (-0.2 in log). The K light would similarly
increase only by $\sim 10\%$ with a similar decrease in $\mlr_K$ (-0.04 in
log). These corrections would have a marginal effect for the conclusions on the
cluster age derived in Figure \ref{fig:bc96}.  Finally the reddening correction
effects on the shape of the stellar light/mass density profiles, would be
negligible for the estimate of the \BH\ mass upper limit.  Indeed, the same
value is obtained regardless of the use of mass density profiles derived from
the I or K band light which are very differently affected by reddening.

Another important issue is the influence of the intrinsic line flux
distribution on \MBH.  We have tried several other different
realizations of the observed flux distribution by varying both the
number of components and the functional form of single components
(exponential, gaussian, power law).
We have found that accurately describing the observed flux distribution is
important. Models that did not do this produced fits significantly worse than those described in the previous section.
When the fits of fluxes and velocities are acceptable, \MBH\ or its
upper limit do not depend on the assumed flux distribution.

The assumption of circular motions is probably one of the most critical ones.
Circular motions in gaseous disks are expected due to the dissipative nature of
the gas, however anisotropies in the stellar potential rapidly lead to
non-circular streaming (\eg, \citealt{athanassoula92}).  The effect of
non-circular motions would be that of ``distorting'' the rotation curves and
increasing the intrinsic line width of the gas.  Therefore, obtaining very good
fits ($\chisqr<1$) of the rotation curves in the alternative model strengthens
our confidence on the assumption of circular motions.  This assumption is also
supported by the small intrinsic line width ($\sigma\simeq 35\kms$) required by
the observations. Indeed $\sigma/\vcirc$ can be used to quantify the
effects of turbulence or non-circular motions on the \BH\ mass determination
(\eg\ \citealt{verdoes00,barth01}).  For the best fitting models $\sigma/\vcirc$
is always less than 0.4 (provided that there is s \BH\ with a mass close to the
upper limit in the case of the alternative models). This is similar to what
\citet{barth01} found in the case of their best fit model of NGC 3245.  There
the effect of the asymmetric drift correction, a possible way to include
non-circular motions in the analysis (\eg\ \citealt{bt87}), is that to increase
the estimate of the \BH\ mass by just $10\%$.  Apart from these qualitative
arguments, the presence of significant non-circular motions cannot be excluded
and it is not possible at the moment to quantify their effects on the method we
have adopted and described here.

Non-circular motions are certainly present in the nuclear region of NGC
4041 as indicated by the presence of the blue wing, but these have
been singled out in the deblending procedure as in 
\citet{winge99}.

\subsection{The blueshift of the nuclear disk}

A proposed explanation for the blueshift observed in the nuclear disk is that
the star cluster is oscillating across the galactic plane. In this picture, the
cluster is bound to the galaxy, is old and has a large \mlr\ ensuring that it
is massive enough and not subjected to tidal disruption.

Any velocity component perpendicular to the disk will give a large
contribution to the observed velocity because the galaxy is almost
face-on. If the cluster is oscillating across the galactic plane, the
observed blueshift may be translated into a velocity modulus of
similar magnitude.  Thus, the cluster velocity ($\sim 10-20\kms$) is
smaller than the rotational velocity (see Figure~\ref{fig:vels}) and the
cluster is bound to the galaxy.  If the cluster is very massive, as in
the alternative model, then the gravitational potential over the star
cluster size is dominated by the star cluster itself and it is not
subjected to tidal disruption. Also in this picture the gaseous disk
is completely dominated by the gravitational potential of the star
cluster.  The total cluster mass within $r<0\farcs3$ as derived from
the I-band mass density profile (the one in which the cluster
dominates) is $\sim 6\xten{7}\Msun$ and this gives an average density
of $\simeq 200\Msun/\PC\3$, not an unusually high value, since it can
be found in galactic globular clusters \citep{globulars}.

In the picture in which the cluster is young (from the fit of the K
band we have $I-K=0.9$, i.e.\ $\sim\ten{7}\YR$ from
Figure~\ref{fig:bc96}), the stability against tidal disruption could
pose a problem, since the mass in the nuclear region is completely
dominated by the bulge stars.  However, this problem could be solved
if the lifetime of the cluster against tidal disruption is less than or
equal to the cluster age.

\subsection{\MBH\  - galaxy correlations}

It is now clear that a large fraction of hot spheroids contain a \BH\ and,
moreover, it seems that the hole mass is proportional to the mass (or
luminosity) of the host spheroid.  Quantitatively, $\MBH/\Msph \approx 0.001$
\citep[\eg][]{merritt01}.  This relation is still controversial, however, both
because the sensitivity of published searches is correlated with bulge
luminosity, and because there is substantial scatter in \MBH\ at fixed \Msph.
Recently \citet{ferrarese00} and \citet{gebhardt00} have shown that a tighter
correlation holds between the BH mass and the velocity dispersion of the bulge.
The two groups however find two different slopes of the correlation,
$\MBH\propto\sigstar^5$ and $\MBH\propto\sigstar^4$ respectively
\citep{tremaine02}.  Clearly, any correlation of black hole and spheroid
properties would have important implications for theories of galaxy formation
in general, and bulge formation in particular.  Indeed, several authors have
shown that the slope of the \MBH-\sigstar\ correlation yields information on
the formation process. If the formation process is self-regulated, i.e.\ if the
BH growth is limited by radiation pressure, $\MBH\propto\sigstar^5$
\citep{sr98}.  Conversely if the growth is regulated by ambient conditions
$\MBH\propto\sigstar^4$ \citep{adams00,cavaliere02}. The uncertainties on the
slope of the \MBH-\sigstar\ correlation do not allow one to distinguish between
the two cases.  To solve this problem more \BH\ mass measurements are needed in
the low mass range (\ten{6}-\ten{7}\Msun) where spiral galaxies are expected to
fit.

These correlations are also very important to estimate \BH\ masses
quickly and easily instead from very complex dynamical and kinematic
measurements.  Therefore the quantities involved in the correlations
must be measured as carefully as possible in order to reduce the
scatter to its intrinsic value.  For instance it has been shown
that with careful estimates of the bulge luminosity the \MBH-\Lsph\
correlation has the same scatter as the \MBH-\sigstar\ correlation, in
contrast with previous claims \citep{mclure02}.

Given the low value of the \BH\ mass in NGC 4041, $\MBH<2\xten{7}\Msun$,
it is worthwhile to verify the relation of this galaxy with the
proposed correlations.  The B magnitude from the RC3 catalogue is 11.9
becoming 11.8 after extinction correction (see NED). The morphological
type is Sbc/Sc and $T=4.0\pm0.3$. From \citet{simien86} the bulge to
total luminosity ratio is $\simeq 0.16$ resulting in $\Delta m = 2$.  Thus the
bulge magnitude is 13.8. The adopted distance, 19.5\MPC, corresponds
to a distance modulus of 31.5 thus the absolute bulge magnitude is
$-17.7$ corresponding to 1.8\xten{9}\LBsun.  The best fit of the
$\MBH-\LBsph$ correlation gives
$\MBH=0.8\xten{8}(\LBsph/\ten{10}\LBsun)^{1.08}$ \citep{kg01}. Thus the
expected \BH\ mass in NGC 4041 would be 1.2\xten{7}\Msun, in agreement
with the \BH\ mass estimate or upper limit.  The best fit of the
$\MBH-\sigstar$ correlation gives
$\MBH=1.3\xten{8}(\sigstar/200\kms)^{4.0}$ \citep{tremaine02} or
$\MBH=1.4\xten{8}(\sigstar/200\kms)^{4.8}$ \citep{merritt01}.
Using INTEGRAL/WYFFOS at the WHT, we have recently measured 
the stellar velocity dispersion in the central 2\arcsec\ of NGC 4041
(Batcheldor et al., in preparation). 
With $\sigstar=95\pm 5\kms$, the
expected \BH\ masses are then 7\xten{6}\Msun\ and 4\xten{6}\Msun, both
consistent with the result from this paper.

Since the main goal of our project is to determine whether or not spirals do in
fact follow the \MBH-\Lsph\ and \MBH-\sigstar\ relations, it is important to
observe also objects which, a-priori, are expected to have marginally or non
detectable \BH s.  Indeed, even if the present measurement is only an upper
limit, this is still useful in ruling out the presence of unusually massive
central \BH s in late type spiral galaxies. 

\section{\label{sec:conclusions}Conclusions}

We presented HST/STIS spectra of the Sbc spiral galaxy NGC 4041 which were used
to map the velocity field of the gas in its nuclear region.  We detected the
presence of a compact ($r\simeq 0\farcs4\simeq 40\PC$), high surface
brightness, circularly rotating nuclear disk cospatial with a nuclear star
cluster. This disk is characterized by a rotation curve with a peak to peak
amplitude of $\sim 40\kms$ and is systematically blueshifted by $\sim 10 -
20\kms$ with respect to the galaxy systemic velocity.

We have analyzed the kinematical data assuming that the stellar mass-to-light
ratio is constant with radius and that the gaseous disk is not warped, having
the same inclination as the large scale galactic disk.  We have found that, in
order to reproduce the observed rotation curve, a dark point mass of
$(1_{-0.7}^{+0.6})\xten{7}\Msun$ is needed, very likely a supermassive \BH.

However the observed blueshift suggests the possibility that the nuclear disk
could be dynamically independent.  Following this line of reasoning we have
relaxed the standard assumptions varying the stellar mass-to-light and the
disk inclination.  We have found that the kinematical data can be accounted for
by the stellar mass provided that either the mass-to-light ratio is increased
by a factor of $\sim 2$ or the inclination is allowed to vary.  This model
resulted in a $3\sigma$ upper limit of $6 \xten{6}\Msun$ on the mass of any
nuclear black hole. 

Combining the results from the standard and alternative models, the present
data only allow us to set an upper limit of $2\xten{7}\Msun$ to the mass of the
nuclear \BH.

If this upper limit is taken in conjunction with an estimated bulge B
magnitude of $-17.7$ and with a central stellar velocity dispersion of $\simeq
95\kms$,
the putative black hole in NGC 4041 is not inconsistent with
both the \MBH-\Lsph\ and the \MBH-\sigstar\ correlations.

\acknowledgments

Support for proposal GO-8228 was provided by NASA through a grant from the
Space Telescope Science Institute, which is operated by the Association of
Universities for Research in Astronomy, Inc., under NASA contract NAS 5-26555

This work was partially supported by the Italian Space Agency (ASI) under
grants I/R/35/00 and I/R/112/01.

This work was partially supported by the Italian Ministry for Instruction,
University and Research (MIUR) under grants Cofin00-02-35 and Cofin01-02-02

We thank Peter Erwin for useful discussions and the referee, Aaron Barth, for 
careful reading of the manuscript, suggestions and comments which improved the
paper.

This publication makes use of the LEDA database (http://leda.univ-lyon1.fr).

This publication makes use of data products from the Two Micron All Sky Survey, which is a joint project of the University of Massachusetts and the Infrared Processing and Analysis Center/California Institute of Technology, funded by the National Aeronautics and Space Administration and the National Science Foundation.

\appendix

\section{\label{app:stardens} Deriving the luminosity density of the stars from
the observed surface brightness profile}

The stellar luminosity density can be inferred by inverting the observed
observed surface brightness profiles.  Following \citet{marel98} we
assume an oblate spheroidal density distribution which we parameterize as:
\begin{equation}\label{eq:massdens}
\rho(m) = \rho_0\left(\frac{m}{r_b}\right)^{-\alpha} \left(1+\left(\frac{m}{r_b}\right)^2\right)^{-\beta}
\end{equation}
$m$ is defined as
$m^2 = x^2+y^2+z^2/q^2$, 
where $xyz$ is a reference system with the $xy$ plane corresponding to the
principal plane of the potential. $q$ is the intrinsic axial ratio of the mass
distribution.  If \mlr\
is the mass-to-light ratio, the observed surface brightness distribution is
given by\,\footnote{Note the $1/4\pi$ factor. This derives from the fact that 
$\rho/V$ is a density (luminosity per unit volume) and $\Sigma$
is a surface brightness (luminosity per unit area per unit solid angle).}:
\begin{equation}
\Sigma = \frac{1}{4\pi\mlr} \int_{-\infty}^{+\infty} \rho ds
\end{equation}
where the integration is performed along the line of sight.  It can be shown
that
\begin{equation}
\Sigma(\ms) = \frac{1}{4\pi\mlr} \frac{\q}{\qs}
\int_{{\ms}^2}^{+\infty} \frac {\rho(m^2) dm^2}{\sqrt{(m^2-\mssq)}}
\end{equation}
where $\mssq = \xssq+\yssq/\qssq$ and \xs\ys\ is a reference system on the sky,
with the \xs\ axis aligned along the apparent major axis. $\qs$ is the observed
axial ratio of the isophotes which is related to the intrinsic
axial ratio of the
mass distribution
by $\qssq = \cos^2\I+q^2 \sin^2\I$, where \I\ is the inclination
of the line of sight ($\I=90$\arcdeg\ is the edge-on case).  The observed surface
brightness results from the convolution of $\Sigma$
with the {\it point spread function} $P$
of the system (i.e.\ telescope and optics) and the detector pixels:
\begin{equation}
\Sigma_\mathrm{app}(X, Y) = \int_{X-\Delta X}^{X+\Delta X} \int_{Y-\Delta Y}^{Y+\Delta Y} \left[ \int_{-\infty}^{+\infty}\int_{-\infty}^{+\infty}\Sigma_\mathrm{true}(x, y) P(\xs-x, \ys-y) \, dx dy \right] 
\,\frac{d\xs d\ys}{4\Delta X\Delta Y}
\end{equation}
where $X,Y$ is the centre of an aperture with size $2\Delta X\times2\Delta Y$.
The integration on $d\xs d\ys$ can be directly carried out on the PSF $P$ which
is described by a sum of gaussians:
\begin{equation}
P(x,y) = \frac{1}{N_K}\sum_{i=1}^N k_i 
\exp\left(-\frac{x^2+y^2}{2\sigma_i^2}\right)
\end{equation}
with $N_K = \sum_{i=1}^N 2\pi\sigma_i^2 k_i$.
Then $\Sigma_\mathrm{app}$ is given by
\begin{equation}
\Sigma_\mathrm{app}(X, Y) = \int_{-\infty}^{+\infty}\int_{-\infty}^{+\infty}
\Sigma(x, y)_\mathrm{true}
\KERN(X-\Delta X-x, X+\Delta X-x;  Y-\Delta Y-y, Y+\Delta Y-y) 
\,\,\frac{dx dy}{4\Delta X\Delta Y}
\end{equation}
where $\KERN$ is the convolution kernel:
\begin{equation}
\KERN(X_1, X_2; Y_1, Y_2) = \frac{1}{N_K}
\sum_{i=1}^N 2\pi\sigma_i^2 \frac{k_i}{4}
\left[\mathcal{E}\left(\frac{X_1}{\sigma_i\sqrt{2}}\right)-\mathcal{E}\left(\frac{X_2}{\sigma_i\sqrt{2}}\right)\right]
\left[\mathcal{E}\left(\frac{Y_1}{\sigma_i\sqrt{2}}\right)-\mathcal{E}\left(\frac{Y_2}{\sigma_i\sqrt{2}}\right)\right]
\end{equation}
where
$\mathcal{E}$ is the complementary error function.
The integration is then carried out numerically using the Gauss Legendre
approximation.

The best fitting model is determined by minimizing the reduced \chisqr\
 defined as:
\begin{eqnarray}
\chisqr & = & \frac{1}{N_d} \sum_{i=1}^{N}  
 \left( \frac{\log\Sigma_{i}-\log\Sigma_m(r_{i}, \delta r_{i}; p_1, \dots, p_m)}{\delta \log\Sigma_{i}} \right)^2  \\
\end{eqnarray}
where $i=1, \dots, N$ indicates a data point with surface brightness $\Sigma_i
\pm \delta\Sigma_{i}$ at radius $r_{i}$. \newline $\Sigma_m(r_{i}, \delta
r_{i}; p_1, \dots, p_m)$ is the model surface brightness, averaged over radii
$r_{i}-\delta r_{i}\le r \le r_{i}+\delta r_{i}$, which is a function of $m$
free parameters $p_1, \dots, p_m$.  $N_d =N - m$ is the number of degrees of
freedom.  The \chisqr\ is minimized to determine the $m$ free parameters using
the downhill simplex algorithm by \citet{numrec}.

When the mass density has been determined as described above,
the circular velocity in the principal plane
is given by the relation \citep[\eg][]{bt87}:
\begin{equation}\label{eq:vcircq}
V_c^2(r) = 4\pi G \q^2 \int_0^r \frac{\rho(m^2) m^2 dm }{\sqrt{r^2-m^2 e^2}}
\end{equation}
where $e$ is the eccentricity related to \q\ by
$\q^2 = 1 -e^2$.
Similarly the mass enclosed within the homoeoid defined by $m<m_\circ$ is
\begin{equation}
M(m_\circ) = 4\pi\q \int_0^{m_\circ} \rho(m^2) m^2 dm
\end{equation}

\section{\label{app:gaskin} The rotation curve model}

\begin{figure}[t!]
\centering
\epsfig{figure=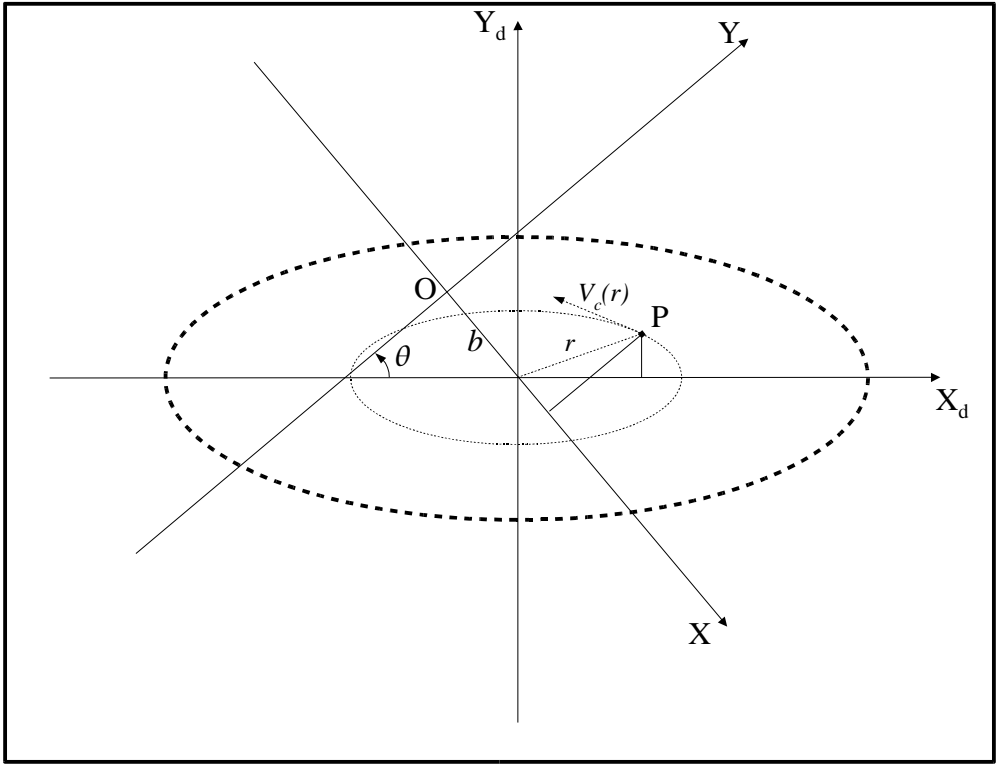,angle=-90,width=0.8\linewidth}
\caption{\label{fig:diskgeo} Geometry of the disk. }
\end{figure}

The velocity field along the line of sight, $\bar{v}$,
can be easily computed in the case of a circularly
rotating thin disk.  Let $XY$ be a reference frame in the plane of the sky
with $Y$ axis fixed along the direction given by the slit position (hereafter called the slit reference frame).
Consider a reference frame $X_dY_d$ in the sky with the $X_d$ axis aligned
along the disk line of nodes and the origin coincident with the disk centre
(hereafter called the ``disk reference frame'',
see Figure~\ref{fig:diskgeo}).  The origin of
$XY$ is chosen in such a way that the disk center has coordinates $x=\B$ and
$y=0$ in the slit reference frame.  Then a given disk point $P$ with
coordinates $(x,y)$ in the slit reference frame has coordinates $(x_d,y_d)$ in
the disk reference frame given by
\begin{eqnarray}
x_d & = & (x-\B) \sin\Th + y \cos\Th \nonumber\\
y_d & = & -(x-\B) \cos\Th + y \sin\Th 
\end{eqnarray}
If the disk has an inclination angle \I\ 
($\I=0$ in the face-on case), then $P$ is at the disk radius $r$ given by
\begin{equation}
r^2 = x_d^2 + \left(\frac{y_d}{\cos\I}\right)^2
\end{equation}
The circular velocity of $P$, in the case of a spherical mass distribution,
is then given by
\begin{equation}
V_c(r) = \left| r\frac{\mathrm{d}\Phi}{\mathrm{d} r} \right|^\frac{1}{2} =
 \left( \frac{GM(r)}{r} \right)^\frac{1}{2}
\end{equation}
where $M(r)$ is the enclosed mass at radius $r$, a constant value \MBH\
in case of a point mass.
In the case of an oblate spheroidal mass distribution, $V_c(r)$ is
given by Equation~\ref{eq:vcircq}.

The velocity component along the line of sight is finally given by
\begin{equation}
\label{eq:kepler}
\bar{v} = \Vsys - V_c (r)\sin \I\frac{x_d}{r} =
\Vsys - (G\MBH)^{0.5}\sin \I\frac{x_d}{r^{1.5}}
\end{equation}
with the latter expression describing the simple case of a point mass \MBH.

This velocity field $\bar{v}$ has to be convolved with the instrumental
response in order to simulate the observed quantities.
Below we show how can this be done for any velocity field $\bar{v}$.

Consider again the reference frame $XY$ in the plane of the sky
defined above. The light distribution can be written as
\begin{equation}\label{eq:w1}
\Phi(\xII,\yII,\vI) = I(\xII,\yII) \phi(\vI-\bar{v}),
\end{equation}
where $I(\xII,\yII)$ is the total intensity at the point $(\xII,\yII)$, and 
$\phi(\vI-\bar{v})$ is the intrinsic line profile centered at the velocity 
along the line of sight $\bar{v}(\xII,\yII)$. 
After passing through the telescope, the light gets convolved with the
instrumental PSF $P(\xI-\xII,\yI-\yII)$, and at point $(\xI,\yI)$ in 
the focal plane the light distribution is described by
\begin{equation}\label{eq:w2}
\Phi(\xI,\yI,\vI) = \int\int_{-\infty}^{+\infty} \dxII\dyII \,\,I(\xII,\yII) 
\phi(\vI-\bar{v}) P(\xI-\xII,\yI-\yII).
\end{equation}
As already defined,
the slit in the focal plane is aligned along the $Y$ axis 
and let its center cross the $X$ axis at $x_0$. Let the slit width be $2\Dx$. 
Past the slit, the light falls into the detector plane,
where the spatial coordinate $\xI$ and the velocity $\vI$ are combined
into one 
single detector coordinate $\wI = \vI + k ( \xI - x_0)$, which we identify
with the {\it observed} velocity (we denote it $w$ here in contrast 
to the intrinsic velocity denoted by $v$). Here, the coefficient $k$ 
is given by $\mu\Dw/\Dy$ where $\mu$, the anamorphic
magnification, accounts for the different scales on the dispersion and slit
directions. In the case of STIS, the scale along the slit is 0\farcs05071 and
along dispersion is 0\farcs05477, thus $\mu = 1.080$ \citep{stisanam}.
The light distribution in the 
detector plane,
$\Psi(\wI, \yI)$, is calculated by ousting $\vI$, and integrating 
the light contribution across the slit
\begin{eqnarray}\label{eq:lprofile0}
\Psi(\wI, \yI) & = &
	\int_{x_0-\Dx}^{x_0+\Dx} \dxI \,\,\Phi(\xI,\yI,\wI - k ( \xI - x_0))
	\nonumber \\
  & = & \int_{x_0-\Dx}^{x_0+\Dx} \dxI 
	\int\int_{-\infty}^{+\infty} \dxII\dyII \,\, 
                           \phi[\wI-\bar{v}(\xII,\yII)-k(\xI-x_0)] \,
			   I(\xII,\yII)\,
			   P(\xI-\xII,\yI-\yII). \nonumber \\
			   & &
\end{eqnarray}
Note the component $k(\xI-x_0)$ in the velocity profile, which is the 
``spurious'' velocity for light entering off from the slit center.
Properties of thus defined two-dimensional light distribution with the spurious
velocity shift were investigated in detail by \citet{witold01}.

The detector integrates over finite pixel sizes therefore
we calculate the expected line fluxes, average velocities and widths
for the line profile that is obtained by integrating the light distribution 
on the detector plane
$\Psi(\wI, \yI)$ over the width $2\Dy$ of the $j^{th}$ pixel 
along the slit, and convolving it with the shape of the pixel in the 
dispersion direction: a top-hat of width $2\Dw$. Thus the expected line profile
in the detector is
\begin{equation}\label{eq:lprofile}
\tilde{\Psi}_j (w) = \int_{y_j-\Dy}^{y_j+\Dy} \dyI  \int_{w-\Dw}^{w+\Dw} 
      d\wI \,\, \Psi(w^\prime, \yI),
\end{equation}
where the $j^{th}$ pixel has the coordinate $y_j$ along the slit. 
Note that the observed intensities are measured at discrete values of $w$
corresponding to the pixel centers.
In order to calculate the expected line fluxes, average 
velocities and widths, one has to evaluate moments of $\tilde{\Psi}_j (w)$:
\begin{equation}\label{eq:mom}
\int_{-\infty}^{+\infty} d\wI \,\, w^n \,\, \tilde{\Psi}_j (w) = 
\int_{y_j-\Dy}^{y_j+\Dy} \dyI \int_{x_0-\Dx}^{x_0+\Dx} \dxI
  \int\int_{-\infty}^{+\infty} \dxII\dyII \,\, \mathcal{P}
  \int_{-\infty}^{+\infty} d\wI \,\, w^n 
  \int_{w-\Dw}^{w+\Dw} \dwp \phi(w'-w_0)
\end{equation}
where we used abbreviations $\mathcal{P} = I(\xII,\yII) P(\xI-\xII,\yI-\yII)$
and $w_0 = \bar{v}(\xII,\yII) + k (x-x_0)$. The last two integrals can be 
simplified by inverting the order of integration:
\[
\int_{-\infty}^{+\infty} d\wI \,\, w^n 
   \int_{w-\Dw}^{w+\Dw} \dwp \phi(w'-w_0) =
\int\int_{-\infty}^{+\infty} d\wI \,\, \dwp \,\, w^n \phi(w'-w_0) H(w-w') = 
\]
\[
\int_{-\infty}^{+\infty} d\wI \,\, \phi(w-w_0) 
   \int_{w-\Dw}^{w+\Dw} \dwp \,\, w'^n =
\int_{-\infty}^{+\infty} d\wI \,\, \phi(w-w_0) 
   \frac{(w+\Dw)^{n+1}-(w-\Dw)^{n+1}}{n+1}
\]
where $H(w-w')$ is 1 for $|w-w'|<\Dw$ and 0 otherwise. This leads to the 
following formulae for the moments of $\tilde{\Psi}_j (w)$:
\begin{equation}\label{eq:intens}
\int_{-\infty}^{+\infty} \tilde{\Psi}_j (w) \dwI = 2\Dw
 	\int_{x_0-\Dx}^{x_0+\Dx} \dxI
	\int_{y_j-\Dy}^{y_j+\Dy} \dyI
	\int\int_{-\infty}^{+\infty} \dxII\dyII \mathcal{P}
\end{equation}
\begin{equation}\label{eq:veloc}
\int_{-\infty}^{+\infty} w \, \tilde{\Psi}_j (w) \dwI = 2\Dw
 	\int_{x_0-\Dx}^{x_0+\Dx} \dxI
	\int_{y_j-\Dy}^{y_j+\Dy} \dyI
	\int\int_{-\infty}^{+\infty} \dxII\dyII w_0 \,\, \mathcal{P}
\end{equation}
\begin{equation}\label{eq:veloc2}
\int_{-\infty}^{+\infty} w^2 \, \tilde{\Psi}_j (w) \dwI = 2\Dw
 	\int_{x_0-\Dx}^{x_0+\Dx} \dxI
	\int_{y_j-\Dy}^{y_j+\Dy} \dyI
	\int\int_{-\infty}^{+\infty} \dxII\dyII 
          (w_0^2+\frac{(\Dw)^2}{3}+\sigma^2)\mathcal{P}
\end{equation}
To obtain them, we assumed that the intrinsic velocity 
profile $\phi(\vI)$ is bound and symmetric, i.e.\ 
\[
\int_{-\infty}^{+\infty} \phi(\vI)\dvI = 1 \hspace{1cm}
\int_{-\infty}^{+\infty} \vI\phi(\vI)\dvI = 0 \hspace{1cm}
\int_{-\infty}^{+\infty} \vI^2\phi(\vI)\dvI = \sigma^2 
\]
where $\sigma^2$ is the intrinsic velocity dispersion. 

Equations~\ref{eq:intens}, \ref{eq:veloc} and \ref{eq:veloc2} can be used to compute
the expected line fluxes, average velocities and widths.
For example, in the most simple case of a constant line intensity
$I=const$, and velocity field $\bar{v}=const$, one can write:
\begin{eqnarray}\label{eq:vel}
\langle v_j\rangle   & = & \frac{\int_{-\infty}^{+\infty} w \, \tilde{\Psi}_j (w) \dwI}
               {\int_{-\infty}^{+\infty} \tilde{\Psi}_j (w) \dwI} = \bar{v} 
	       \nonumber \\
\langle v_j^2\rangle & = & \frac{\int_{-\infty}^{+\infty} w^2 \, \tilde{\Psi}_j (w) \dw}
               {\int_{-\infty}^{+\infty} \tilde{\Psi}_j (w) \dw} =
		 \bar{v}^2 + \sigma^2 + \frac{(\Dw)^2}{3} + \frac{(k\Dx)^2}{3}.
\end{eqnarray}
Then the expected velocity dispersion, given by 
$\sigma_j^2 = \langle v_j^2 \rangle- \langle v_j \rangle ^2$, is,
understandably, larger
than the intrinsic velocity dispersion $\sigma$. This is due to the convolution with the pixel
size and slit width, which add quadratically. However, note that while $\Dw$ and
$\sigma$ enter the integral (\ref{eq:veloc2}) as constants, $w_0$ is a linear
combination of the spurious velocity shift, and the intrinsic velocity. These
two contributions can cancel, resulting in the expected line profile being broadened
by the pixel size only, and not by the width of the slit. This implies that
the wide slit can probe the disk on the scale of the pixel size rather than the
slit width. \citet{witold01} explore consequences of this finding.

In order to compute the model given by
Equations \ref{eq:intens}, \ref{eq:veloc}, 
and \ref{eq:veloc2}, we create a grid in $x$ and $y$ with sampling given by 
$\sigma_\mathrm{PSF}/n$. Here, $\sigma_\mathrm{PSF}$ is the spatial r.m.s. 
of the point spread function (PSF) and $n$ is the subsampling factor. We 
have verified that the optimal subsampling factor used is $n=3$ since larger 
values do not produce any appreciable differences in the final results. The 
PSF used is the one generated by TinyTim (V6.0, \citealt{tinytim}) at 6700\AA. 
Convolution with the PSF is done using the {\it Fast Fourier Transform} 
algorithm \citet{numrec}.
Following \citet{barth01} we have also introduced the CCD scattering
function \citet{stishand} but, as already noticed by them, it does
not have any appreciable effect in the final results.

We compare models to the observed spectrum, which is essentially  
an array of intensities $\Psi_{ij}$, after the observed line profile
is derived in the following way: to the sequence of intensities $\Psi_{ij}$
for a given row $j$ along the slit, we fit a baseline, and a continuous 
analytical function $\tilde{\Psi}_j(w)^{\rm obs}$, which we interpret as 
the observed equivalent of the expected line profile $\tilde{\Psi}_j(w)$ 
(eq.\ref{eq:lprofile}).
The best fitting model is determined by minimizing the reduced \chisqr\
 defined as:
\begin{equation}
\chisqr =  \frac{1}{N_d} \sum_{k=1}^{3} \sum_{j=1}^{N_k} \left[ 
 \left( \frac{v_{kj}-\langle v_j \rangle _k (p_1, \dots, p_m)}{\delta v_{kj}} \right)^2 + 
 \left( \frac{W_{kj}-\langle W_j \rangle _k (p_1, \dots, p_m)}{\delta W_{kj}} \right)^2 \right]
\end{equation}
where the index $k=1,3$ indicates the slit position, and $j=1,N_k$ counts 
pixels along the slit. Here, the characteristics of the model are as follows. The velocity 
in the $j^{th}$ row along the slit $\langle v_j \rangle _k$, and the FWHM 
of the velocity profile $\langle W_j \rangle _k$, are calculated directly
from equations (\ref{eq:vel}), now clearly for variable 
intensity and velocity field. They both are functions of $m$ free parameters 
$p_1, \dots, p_m$, which are determined by $\chisqr$ minimization.
The FWHM is calculated from the expected 
velocity dispersion $\sigma_j$ after assuming a Gaussian line profile.
The observed velocities ($v_{ki}\pm \delta v_{ki}$), and velocity
dispersions ($W_{ki}\pm \delta W_{ki}$) are also derived from equations 
(\ref{eq:vel}), but after $\tilde{\Psi}_j(w)$ has been
replaced by $\tilde{\Psi}_j(w)^{\rm obs}$ defined above.
$N_d =\sum_{k=1}^{3} 2N_k - m$ is the number of degrees of freedom.

The \chisqr\ is minimized to determine the $m$ free parameters using the
downhill simplex algorithm by \citet{numrec}. In order to apply statistical
methods when \chisqr\ is much larger than 1, we follow \citet{barth01} and
rescale errors as:
\begin{equation}
(\delta v_{ki}^\prime)^2 = \delta v_{ki}^2 + \delta V^2
\end{equation} 
where $\delta V$ is a ``systematic'' error determined such that the resulting
\chisqc\ of the ``best'' model is 1. The fits presented in this
paper have $\chisqr\sim 1$ and the error rescaling was not performed.

\end{document}